\documentstyle[prd,aps,preprint,tighten,epsfig]{revtex}

\begin{document}

\draft

\title{Implications of the Daya Bay observation of $\theta^{}_{13}$ on
the leptonic flavor mixing structure and CP violation}
\author{{\bf Zhi-zhong Xing}
\thanks{E-mail: xingzz@ihep.ac.cn}}
\address{Institute of High Energy Physics, Chinese Academy of Sciences,
Beijing 100049, China}

\maketitle

\begin{abstract}
The Daya Bay Collaboration has recently reported its first
$\overline{\nu}^{}_e \to \overline{\nu}^{}_e$ oscillation result
which points to $\theta^{}_{13} \simeq 8.8^\circ \pm 0.8^\circ$
(best-fit $\pm 1\sigma$ range) or $\theta^{}_{13} \neq 0^\circ$ at
the $5.2\sigma$ level. The fact that this smallest neutrino mixing
angle is not strongly suppressed motivates us to look into the
underlying structure of lepton flavor mixing and CP violation. Two
phenomenological strategies are outlined: (1) the lepton flavor
mixing matrix $U$ consists of a constant leading term $U^{}_0$ and a
small perturbation term $\Delta U$; and (2) the mixing angles of $U$
are associated with the lepton mass ratios. Some typical patterns of
$U^{}_0$ are reexamined by constraining their respective
perturbations with current experimental data. We illustrate a few
possible ways to minimally correct $U^{}_0$ in order to fit the
observed values of three mixing angles. We point out that the
structure of $U$ may exhibit an approximate $\mu$-$\tau$
permutation symmetry in modulus,
and reiterate the geometrical description of
CP violation in terms of the leptonic unitarity triangles. The
salient features of nine distinct parametrizations of $U$ are
summarized, and its Wolfenstein-like expansion is presented by
taking $U^{}_0$ to be the democratic mixing pattern.
\end{abstract}


\newpage

\section{Introduction}

Thanks to a number of well-done solar, atmospheric, reactor and
accelerator neutrino oscillation experiments, we are now convinced
that three known neutrinos have finite masses and one lepton flavor
can convert to another \cite{PDG10}. The phenomenon of lepton flavor
mixing at low energies is effectively described by a $3\times 3$
matrix $U$, the so-called Maki-Nakagawa-Sakata-Pontecorvo (MNSP)
matrix \cite{MNS}, in the weak charged-current interactions:
\begin{eqnarray}
-{\cal L}^{}_{\rm cc} = \frac{g}{\sqrt{2}} \ \overline{
\left(\matrix{e & \mu & \tau} \right)^{}_{\rm L}} \ \gamma^\mu
\left( \matrix{ U^{}_{e1} & U^{}_{e2} & U^{}_{e3} \cr U^{}_{\mu 1} &
U^{}_{\mu 2} & U^{}_{\mu 3} \cr U^{}_{\tau 1} & U^{}_{\tau 2} &
U^{}_{\tau 3} \cr} \right) \left(\matrix{ \nu^{}_1 \cr \nu^{}_2 \cr
\nu^{}_3 \cr} \right)^{}_{\rm L} W^-_\mu + {\rm h.c.} \; .
\end{eqnarray}
Given the unitarity of $U$, it can be parametrized in terms of three
angles and three phases:
\begin{eqnarray}
U = \left( \matrix{ c^{}_{12}
c^{}_{13} & s^{}_{12} c^{}_{13} & s^{}_{13} e^{-i\delta} \cr
-s^{}_{12} c^{}_{23} - c^{}_{12} s^{}_{13} s^{}_{23} e^{i\delta} &
c^{}_{12} c^{}_{23} - s^{}_{12} s^{}_{13} s^{}_{23} e^{i\delta} &
c^{}_{13} s^{}_{23} \cr s^{}_{12} s^{}_{23} - c^{}_{12} s^{}_{13}
c^{}_{23} e^{i\delta} & -c^{}_{12} s^{}_{23} - s^{}_{12} s^{}_{13}
c^{}_{23} e^{i\delta} & c^{}_{13} c^{}_{23} \cr} \right) P^{}_\nu \;
,
\end{eqnarray}
where $c^{}_{ij} \equiv \cos\theta^{}_{ij}$, $s^{}_{ij} \equiv
\sin\theta^{}_{ij}$ (for $ij = 12, 13, 23$), and $P^{}_\nu ={\rm
Diag}\{e^{i\rho}, e^{i\sigma}, 1\}$ which is physically relevant if
massive neutrinos are the Majorana particles. A global analysis of
the available neutrino oscillation data \cite{FIT} points to
$\theta^{}_{12} \simeq 34^\circ$ and $\theta^{}_{23} \simeq
45^\circ$, much larger than the Cabibbo angle $\vartheta^{}_{\rm C}
\simeq 13^\circ$ in the Cabibbo-Kobayashi-Maskawa (CKM) quark flavor
mixing matrix $V$ \cite{CKM}. The third mixing angle
$\theta^{}_{13}$ is expected to be smaller than $\vartheta^{}_{\rm
C}$, and its central value might be around $8^\circ$ \cite{Minakata}
as hinted by the preliminary T2K \cite{T2K}, MINOS \cite{MINOS} and
Double Chooz \cite{DC} data. Three CP-violating phases of $U$ remain
unknown at this stage, but one of them (i.e., the Dirac phase
$\delta$) will be measured in the forthcoming long-baseline neutrino
oscillation experiments.

The Daya Bay Collaboration has recently made a breakthrough
in the measurement of $\theta^{}_{13}$
from the reactor $\overline{\nu}^{}_e \to \overline{\nu}^{}_e$
oscillations \cite{DYB}. The best-fit ($\pm 1\sigma$ range) result is
\begin{eqnarray}
\sin^2 2\theta^{}_{13} = 0.092 \pm 0.016 ({\rm stat})
\pm 0.005 ({\rm syst}) \; ,
\end{eqnarray}
which is equivalent to $\theta^{}_{13} \simeq 8.8^\circ \pm
0.8^\circ$ or $\theta^{}_{13} \neq 0^\circ$ at the $5.2\sigma$
level. This very encouraging observation convinces us that the
smallest neutrino mixing angle is not really small and the MNSP
matrix $U$ is not strongly hierarchical. We are therefore motivated
to study the underlying structure of lepton flavor mixing and CP
violation. In fact, $U$ has been conjectured to have the following
structure for a quite long time \cite{FX96}:
\begin{eqnarray}
U = \left(U^{}_0 + \Delta U\right) P^{}_\nu \; ,
\end{eqnarray}
in which the leading term $U^{}_0$ is a constant matrix responsible
for two larger mixing angles $\theta^{}_{12}$ and $\theta^{}_{23}$,
and the next-to-leading term $\Delta U$ is a perturbation matrix
responsible for both the smallest mixing angle $\theta^{}_{13}$ and
the Dirac CP-violating phase $\delta$. So far a lot of flavor
symmetries have been brought into exercise to derive $U^{}_0$, while
$\Delta U$ might originate from either an explicit flavor symmetry
breaking scenario or some finite quantum corrections at a given
energy scale or from a superhigh-energy scale to the electroweak
scale. In view of the new and robust Daya Bay result for
$\theta^{}_{13}$, we are immediately concerned about two burning
issues of the day in the phenomenology of neutrino physics:
\begin{itemize}
\item     If the essential structure of lepton flavor mixing is really
revealed by Eq. (4), can there be a natural pattern of
$U^{}_0$ accompanied by a natural perturbation matrix $\Delta U$?

\item     If the main part of the MNSP matrix $U$ is not a constant
mixing matrix, what is the most straightforward way to understand
the salient features of lepton flavor mixing?
\end{itemize}
In addition, we are curious about whether the structure of $U$ has
an approximate $\mu$-$\tau$ permutation symmetry in modulus,
whether leptonic
CP violation is significant in neutrino oscillations, whether the
other parametrizations of $U$ besides the one in Eq. (2) are useful
for describing the properties of lepton flavor mixing and CP
violation, and whether there is an interesting and suggestive
expansion of $U$ as compared with the popular Wolfenstein
parametrization of the CKM matrix $V$ \cite{W}, and so on.

The purpose of this paper is to answer those easy questions and
outline some possible ways to deal with those difficult ones as
mentioned above. In section II we describe two phenomenological
strategies towards understanding the textures of lepton mass
matrices and thus the structure of lepton flavor mixing: one of them
can result in Eq. (4), and the other is expected to relate the
mixing angles to the lepton mass ratios. In section III we reexamine
five typical patterns of $U^{}_0$ (the democratic \cite{FX96},
bimaximal \cite{BM}, tri-bimaximal \cite{TB}, golden-ratio \cite{GR}
and hexagonal \cite{Xing03} mixing patterns) by estimating their
respective perturbation matrices with the help of the latest Daya
Bay result for $\theta^{}_{13}$. Except the democratic mixing
pattern, we find that the other four patterns of $U^{}_0$ suffer
from a common problem: the viable perturbation matrix $\Delta U$ has
to be adjusted in a more or less unnatural way to make one or two of
the large mixing angles of $U^{}_0$ slightly modified but its
smallest (vanishing) angle significantly corrected. Section IV is
devoted to a brief discussion about the possible minimal
perturbations to $U^{}_0$. We take three interesting examples to
illustrate three simple approaches for this goal. In section V we
point out the conditions under which the MNSP matrix $U$ may have an
exact or approximate $\mu$-$\tau$ permutation symmetry in modulus.
The strength
of leptonic CP violation is calculated, and the language of leptonic
unitarity triangles is reiterated to geometrically describe CP
violation. Section VI is devoted to a summary of nine topologically
distinct parametrizations of $U$ and their respective features or
merits, and section VII is devoted to a Wolfenstein-like expansion
of $U$ by taking $U^{}_0$ to be the democratic mixing pattern. In
section VIII we first summarize the main points and results of this
paper and then make some concluding remarks.

\section{Two phenomenological strategies}

The MNSP matrix $U$ actually describes a fundamental mismatch
between the weak-interaction (flavor) and mass eigenstates of six
leptons, or equivalently a mismatch between the diagonalizations of
the charged-lepton mass matrix $M^{}_l$ and the effective neutrino
mass matrix $M^{}_\nu$ in a given model, no matter whether the
origin of neutrino masses is attributed to the seesaw mechanisms or
not \cite{FX00}. Assuming massive neutrinos to be the Majorana
particles, we may simply write out the leptonic mass terms as
\begin{eqnarray}
-{\cal L}^{}_{\rm mass} = \overline{
\left(\matrix{e^\prime & \mu^\prime & \tau^\prime} \right)^{}_{\rm L}}
\ M^{}_l \left(\matrix{ e^\prime \cr \mu^\prime \cr \tau^\prime}
\right)^{}_{\rm R} + \ \frac{1}{2} \
\overline{ \left(\matrix{\nu^{}_e & \nu^{}_\mu & \nu^{}_\tau}
\right)^{}_{\rm L}} \ M^{}_\nu
\left(\matrix{ \nu^{c}_e \cr \nu^{c}_\mu \cr
\nu^{c}_\tau \cr} \right)^{}_{\rm R} + {\rm h.c.} \; ,
\end{eqnarray}
in which ``$\prime$" stands for the flavor eigenstates of charged
leptons, ``$c$" denotes the charge-conjugated neutrino fields, and
$M^{}_\nu$ is symmetric. By means of the unitary matrices $O^{}_l$,
$O^\prime_l$ and $O^{}_\nu$, one can diagonalize $M^{}_l$ and
$M^{}_\nu$ through the transformations $O^\dagger_l M^{}_l O^\prime_l =
\widehat{M}^{}_l \equiv {\rm Diag}\{m^{}_e, m^{}_\mu, m^{}_\tau\}$
and $O^\dagger_\nu M^{}_\nu O^*_\nu = \widehat{M}^{}_\nu \equiv {\rm
Diag}\{m^{}_1, m^{}_2, m^{}_3 \}$, respectively. Then one arrives at
the lepton mass terms in terms of the mass eigenstates:
\begin{eqnarray}
-{\cal L}^\prime_{\rm mass} = \overline{
\left(\matrix{e & \mu & \tau} \right)^{}_{\rm L}}
\ \widehat{M}^{}_l \left(\matrix{ e \cr \mu \cr \tau}
\right)^{}_{\rm R} + \ \frac{1}{2} \
\overline{ \left(\matrix{\nu^{}_1 & \nu^{}_2 & \nu^{}_3}
\right)^{}_{\rm L}} \ \widehat{M}^{}_\nu
\left(\matrix{ \nu^{c}_1 \cr \nu^{c}_2 \cr
\nu^{c}_3 \cr} \right)^{}_{\rm R} + {\rm h.c.} \; .
\end{eqnarray}
Extending this basis transformation to the standard weak
charged-current interactions, we immediately obtain Eq. (1) in which
the MNSP matrix $U$ is given by $U = O^\dagger_l O^{}_\nu$.

The above treatment is most general at a given energy scale
(e.g., the electroweak scale), but it can still provide
us with the following lessons:
\begin{itemize}
\item     The structure of lepton flavor mixing is directly
determined by the structures of $O^{}_l$ and $O^{}_\nu$. Since these
two unitary matrices are used to diagonalize $M^{}_l$ and $M^{}_\nu$,
respectively, their structures are governed by those of $M^{}_l$
and $M^{}_\nu$, whose eigenvalues are the physical lepton masses.
Therefore, we anticipate that the dimensionless flavor mixing angles
of $U$ should be certain kinds of functions whose variables include
four independent mass ratios of three charged leptons and three
neutrinos. Namely,
\begin{eqnarray}
\theta^{}_{ij} = f\left(\frac{m^{}_\alpha}{m^{}_\beta},
\frac{m^{}_k}{m^{}_l}, \cdots\right) \; ,
\end{eqnarray}
where the Greek subscripts denote the charged leptons, the Latin
subscripts stand for the neutrinos, and ``$\cdots$" implies other
possible dimensionless parameters originating from the lepton mass
matrices. Such an expectation has proved valid in the quark sector
to explain why the relation $\sin\vartheta^{}_{\rm C} \simeq
\sqrt{m^{}_d/m^{}_s}$ works quite well and how the hierarchical
structure of the CKM matrix $V$ is related to the strong hierarchies
of quark masses (i.e., $m^{}_u \ll m^{}_c \ll m^{}_t$ and $m^{}_d
\ll m^{}_s \ll m^{}_b$) \cite{FX00}. As for the phenomenon of lepton
flavor mixing, it is apparently difficult to link two large mixing
angles $\theta^{}_{12}$ and $\theta^{}_{23}$ to two small mass
ratios $m^{}_e/m^{}_\mu \simeq 4.7 \times 10^{-3}$ and
$m^{}_\mu/m^{}_\tau \simeq 5.9 \times 10^{-2}$ \cite{XZZ}. Hence one
may consider to ascribe the largeness of $\theta^{}_{12}$ and
$\theta^{}_{23}$ to a weak hierarchy of three neutrino
masses, such as the conjecture
$\tan\theta^{}_{12} \simeq \sqrt{m^{}_1/m^{}_2}$ \cite{FX06} or
$\sin\theta^{}_{13} \simeq \sin\theta^{}_{12} \sqrt{m^{}_2/m^{}_3}$
\cite{R}.

\item     To establish a direct relation between $\theta^{}_{ij}$
and lepton mass ratios, one has to specify the textures of $M^{}_l$
and $M^{}_\nu$ by allowing some of their elements to vanish or to
be vanishingly small. The
most instructive example of this kind is the Fritzsch ansatz
\cite{Fritzsch},
\begin{eqnarray}
M^{}_{l,\nu} = \left( \matrix{ 0 & \times & 0 \cr
\times & 0 & \times \cr 0 & \times & \times \cr} \right) \; ,
\end{eqnarray}
which is able to account for current neutrino oscillation data to an
acceptable degree of accuracy (e.g., $\sin\theta^{}_{23} \simeq
\sqrt{m^{}_\mu/m^{}_\tau} + \sqrt{m^{}_2/m^{}_3} \simeq 0.65$)
\cite{Xing02}. Another well-known and phenomenologically viable
example is the two-zero textures of $M^{}_\nu$ in the basis where
$M^{}_l$ is diagonal \cite{Zero}. Note that the texture zeros of a
fermion mass matrix dynamically mean that the corresponding matrix
elements are sufficiently suppressed as compared with their
neighboring counterparts, and they can be derived from a certain
flavor symmetry in a given theoretical framework (e.g., with the
help of the Froggatt-Nielson mechanism \cite{FN} or discrete flavor
symmetries \cite{Grimus}).

\item     We realize that the expectation in Eq. (7)
is actually in conflict with the conjecture made in Eq. (4). In
other words, the leading term of the MNSP matrix $U$ might be a
constant matrix whose mixing angles are independent of the lepton
mass ratios. The reason for this ``conflict" is rather simple: the
assumed structures of lepton flavor mixing in Eqs. (4) and (7)
correspond to two different structures of lepton mass matrices. As
we have pointed out above, the direct dependence of $\theta^{}_{ij}$
on $m^{}_\alpha/m^{}_\beta$ and $m^{}_k/m^{}_l$ is usually a direct
consequence of the texture zeros of $M^{}_l$ and (or) $M^{}_\nu$. In
contrast, a constant flavor mixing pattern $U^{}_0$ may arise from
some special textures of $M^{}_l$ and (or) $M^{}_\nu$ whose entries
have certain kinds of linear correlations or equalities. For
instance, the texture \cite{FL}
\begin{eqnarray}
M^{}_\nu = \left( \matrix{ b+c & -b & -c \cr -b & a+b & -a \cr
-c & -a & a+c \cr} \right) \;
\end{eqnarray}
assures $O^{}_\nu$ to be of the tri-bimaximal mixing pattern. This
neutrino mass matrix has no zero entries, but its nine elements
satisfy the sum rules $(M^{}_\nu)^{}_{i1} + (M^{}_\nu)^{}_{i2} +
(M^{}_\nu)^{}_{i3} = 0$ and $(M^{}_\nu)^{}_{1j} + (M^{}_\nu)^{}_{2j}
+ (M^{}_\nu)^{}_{3j} = 0$ (for $i, j=1,2,3$). Such correlative
relations are similar to those texture zeros in the sense that both
of them may reduce the number of free parameters associated with
lepton mass matrices, making some predictions for the lepton flavor
mixing angles technically possible.

\item     It is well known that the special textures of $M^{}_l$
and $M^{}_\nu$ like that in Eq. (9) can easily be derived from
certain discrete flavor symmetries (e.g., $A^{}_4$ or $S^{}_4$)
\cite{Flavor}. That is why Eq. (4) formally summarizes a large class
of lepton flavor mixing patterns in which the leading terms are
constant matrices originating from some underlying flavor
symmetries. The fact that $\theta^{}_{13}$ is not very small poses a
meaningful question to us today: can this mixing angle naturally be
generated from the perturbation matrix $\Delta U$? The answer to
this question is certainly dependent upon the form of $U^{}_0$ in
the flavor symmetry limit. We shall reexamine five typical patterns of
$U^{}_0$ in the subsequent section to get a feeling of the
respective structures of $\Delta U$ which can be constrained by current
experimental data on neutrino oscillations.
\end{itemize}

In short, one may try to understand the structure of the MNSP matrix $U$
by following two phenomenological strategies: one is to explore
possible relations between the flavor mixing angles and the lepton
mass ratios, and the other is to investigate possible constant
patterns of lepton flavor mixing as the leading-order effects. We
have seen that the former possibility essentially points to some
vanishing (or vanishingly small) entries of $M^{}_l$ and $M^{}_\nu$,
while the latter possibility apparently indicates some equalities or
linear correlations among the entries of $M^{}_l$ or $M^{}_\nu$. In
both cases the underlying flavor symmetries play a crucial role in
deriving the structures of lepton mass matrices which finally
determine the structure of lepton flavor mixing. Of course, how to
pin down the correct flavor symmetries remains an open question.

\section{Five patterns of the MNSP matrix}

For the sake of simplicity, we typically take $\theta^{}_{12} \simeq
34^\circ$, $\theta^{}_{13} \simeq 9^\circ$ and $\theta^{}_{23}
\simeq 45^\circ$ as our inputs to fix the primary structure of the
MNSP matrix $U$. Then we arrive at
\begin{eqnarray}
U = \left( \matrix{ 0.819 & 0.552 & 0.156 e^{-i\delta} \cr
-0.395 - 0.092 e^{i\delta} &
0.586 - 0.062 e^{i\delta} &
0.698 \cr 0.395 - 0.092 e^{i\delta} & -0.586 - 0.062
e^{i\delta} & 0.698 \cr} \right) P^{}_\nu \; .
\end{eqnarray}
It makes sense to compare a constant mixing pattern $U^{}_0$ with
the observed pattern of $U$ in Eq. (10), such that one may estimate
the structure of the corresponding perturbation matrix $\Delta U$.
Let us consider five well-known patterns of $U^{}_0$ for
illustration.

(1) The democratic mixing pattern of lepton flavors \cite{FX96}:
\begin{eqnarray}
U^{}_0 = \left( \matrix{ \frac{1}{\sqrt 2} & \frac{1}{\sqrt 2} & 0
\cr -\frac{1}{\sqrt 6} & \frac{1}{\sqrt 6} & \frac{\sqrt 2}{\sqrt 3}
\cr \frac{1}{\sqrt 3} & -\frac{1}{\sqrt 3} & \frac{1}{\sqrt 3} \cr}
\right) \; ,
\end{eqnarray}
whose three mixing angles are $\theta^{(0)}_{12} = 45^\circ$,
$\theta^{(0)}_{13} = 0^\circ$ and $\theta^{(0)}_{23}
=\arctan(\sqrt{2}) \simeq 54.7^\circ$ in the standard
parametrization as given in Eq. (2). With the help of Eq. (10), we
immediately obtain the form of $\Delta U = UP^\dagger_\nu - U^{}_0$
as follows:
\begin{eqnarray}
\Delta U = \left( \matrix{ 0.112
& -0.155 & 0.156 e^{-i\delta} \cr
0.013 - 0.092 e^{i\delta} &
0.178 - 0.062 e^{i\delta} &
-0.118 \cr -0.182 - 0.092 e^{i\delta} & -0.009 - 0.062
e^{i\delta} & 0.121 \cr} \right) \; .
\end{eqnarray}
One can see that the magnitude of each matrix element of $\Delta U$
is of ${\cal O}(0.1)$, implying that the realistic pattern of $U$
might result from a democratic perturbation to $U^{}_0$ (i.e., the
nine entries of $\Delta U$ are all proportional to a common small
parameter). We shall elaborate on this point in
detail in section VII.

(2) The bimaximal mixing pattern of lepton flavors \cite{BM}:
\begin{eqnarray}
U^{}_0 = \left( \matrix{ \frac{1}{\sqrt 2} & \frac{1}{\sqrt 2} & 0
\cr -\frac{1}{2} & \frac{1}{2} & \frac{1}{\sqrt 2} \cr \frac{1}{2} &
-\frac{1}{2} & \frac{1}{\sqrt 2} \cr} \right) \; ,
\end{eqnarray}
which has $\theta^{(0)}_{12} = 45^\circ$, $\theta^{(0)}_{13} =
0^\circ$ and $\theta^{(0)}_{23} = 45^\circ$ in the standard
parametrization. Comparing Eq. (13) with Eq. (10), we obtain the
perturbation matrix
\begin{eqnarray}
\Delta U = \left( \matrix{ 0.112
& -0.155 & 0.156 e^{-i\delta} \cr
0.105 - 0.092 e^{i\delta} &
0.086 - 0.062 e^{i\delta} &
-0.009 \cr -0.105 - 0.092 e^{i\delta} & -0.086 - 0.062
e^{i\delta} & -0.009 \cr} \right) \; .
\end{eqnarray}
We see that the matrix elements $(\Delta U)^{}_{\mu 3}$ and $(\Delta
U)^{}_{\tau 3}$ are highly suppressed. In other words, the initially
maximal angle $\theta^{(0)}_{23}$ receives the minimal correction,
which is much smaller than the one received by the initially minimal
angle $\theta^{(0)}_{13}$. Such a situation seems to be more or
less unnatural, at least from a point of view of model building.

(3) The tri-bimaximal mixing pattern of lepton flavors
\cite{TB}:
\begin{eqnarray}
U^{}_0 = \left( \matrix{ \frac{\sqrt 2}{\sqrt 3} & \frac{1}{\sqrt 3}
& 0 \cr -\frac{1}{\sqrt 6} & \frac{1}{\sqrt 3} & \frac{1}{\sqrt 2}
\cr \frac{1}{\sqrt 6} & -\frac{1}{\sqrt 3} & \frac{1}{\sqrt 2} \cr}
\right) \; ,
\end{eqnarray}
whose three mixing angles are $\theta^{(0)}_{12} =
\arctan(1/\sqrt{2}) \simeq 35.3^\circ$, $\theta^{(0)}_{13} =
0^\circ$ and $\theta^{(0)}_{23} =45^\circ$ in the standard
parametrization. In a similar way we get the corresponding
perturbation matrix
\begin{eqnarray}
\Delta U = \left( \matrix{ 0.003
& -0.025 & 0.156 e^{-i\delta} \cr
0.013 - 0.092 e^{i\delta} &
0.009 - 0.062 e^{i\delta} &
-0.009 \cr -0.013 - 0.092 e^{i\delta} & -0.009 - 0.062
e^{i\delta} & -0.009 \cr} \right) \; .
\end{eqnarray}
It is quite obvious that $(\Delta U)^{}_{e 1}$, $(\Delta U)^{}_{e 2}$,
$(\Delta U)^{}_{\mu 3}$ and $(\Delta U)^{}_{\tau 3}$ are highly
suppressed. So two initially large angles $\theta^{(0)}_{12}$ and
$\theta^{(0)}_{23}$ are only slightly modified by the perturbation
effects, but the initially minimal angle $\theta^{(0)}_{13}$ receives
the maximal correction.

(4) The golden-ratio mixing pattern of lepton flavors
\cite{GR}:
\begin{eqnarray}
U^{}_0 = \left( \matrix{ \frac{\sqrt 2}{\sqrt{5 - \sqrt 5}}
& \frac{\sqrt 2}{\sqrt{5 + \sqrt 5}} & 0
\cr -\frac{1}{\sqrt{5 + \sqrt 5}} & \frac{1}{\sqrt{5 - \sqrt 5}}
& \frac{1}{\sqrt 2} \cr \frac{1}{\sqrt{5 + \sqrt 5}}
& -\frac{1}{\sqrt{5 - \sqrt 5}} & \frac{1}{\sqrt 2} \cr}
\right) \; ,
\end{eqnarray}
which has $\theta^{(0)}_{12} = \arctan[2/(1+ \sqrt{5})] \simeq
31.7^\circ$, $\theta^{(0)}_{13} = 0^\circ$ and $\theta^{(0)}_{23}
=45^\circ$ in the standard parametrization. In this case the
perturbation matrix $\Delta U$ turns out to be
\begin{eqnarray}
\Delta U = \left( \matrix{ -0.032
& 0.026 & 0.156 e^{-i\delta} \cr
-0.023 - 0.092 e^{i\delta} &
-0.016 - 0.062 e^{i\delta} &
-0.009 \cr 0.023 - 0.092 e^{i\delta} & 0.016 - 0.062
e^{i\delta} & -0.009 \cr} \right) \; .
\end{eqnarray}
Similar to the tri-bimaximal mixing pattern, two initially large
angles of the golden-ratio mixing pattern are only slightly
corrected, but the initially minimal angle $\theta^{(0)}_{13}$ is
significantly modified by the same perturbation.

(5) The hexagonal mixing pattern of lepton flavors
\cite{Xing03}:
\begin{eqnarray}
U^{}_0 = \left( \matrix{ \frac{\sqrt 3}{2} & \frac{1}{2} & 0 \cr
-\frac{\sqrt 2}{4} & \frac{\sqrt 6}{4} & \frac{1}{\sqrt 2} \cr
\frac{\sqrt 2}{4} & -\frac{\sqrt 6}{4} & \frac{1}{\sqrt 2} \cr}
\right) \; ,
\end{eqnarray}
whose mixing angles are $\theta^{(0)}_{12} =
30^\circ$, $\theta^{(0)}_{13} = 0^\circ$ and $\theta^{(0)}_{23}
=45^\circ$ in the standard parametrization. In this case we obtain
the perturbation matrix
\begin{eqnarray}
\Delta U = \left( \matrix{ -0.047 & 0.052 & 0.156 e^{-i\delta} \cr
-0.041 - 0.092 e^{i\delta} & -0.026 - 0.062 e^{i\delta} & -0.009 \cr
0.041 - 0.092 e^{i\delta} & 0.026 - 0.062 e^{i\delta} & -0.009 \cr}
\right) \; .
\end{eqnarray}
This result is quite analogous to the one obtained in Eq. (16) or
Eq. (18), simply because the patterns of $U^{}_0$ in these three
cases are quite similar.

Now let us summarize some useful lessons that we can directly
learn from the above five typical examples of $U$.
\begin{itemize}
\item     To accommodate the unsuppressed value of
$\theta^{}_{13}$ in a generic flavor mixing structure $U =
\left(U^{}_0 + \Delta U\right) P^{}_\nu$, one has to choose a
proper constant mixing pattern $U^{}_0$ and adjust its perturbation
matrix $\Delta U$. The phenomenological criterion to do so is
two-fold: on the one hand, $U^{}_0$ should easily be derived from a
certain flavor symmetry; on the other hand, $\Delta U$ should have a
natural structure which can easily be accounted for by either the
flavor symmetry breaking or quantum corrections (or both of them).

\item     The common feature of the above five patterns of
$U^{}_0$ is apparently $(U^{}_0)^{}_{e 3} = 0$ (or equivalently,
$\theta^{(0)}_{13} =0^\circ$), implying that a relatively large
perturbation is required for generating $\theta^{}_{13} \sim
9^\circ$. In this case, the closer $\theta^{(0)}_{12}$ and
$\theta^{(0)}_{23}$ are to the observed values of $\theta^{}_{12}$
and $\theta^{}_{23}$, the more unnatural the structure of $\Delta U$
seems to be. The tri-bimaximal mixing pattern given in Eq. (15),
which is currently the most popular ansatz for model building based
on certain flavor symmetries, suffers from this unnaturalness in
particular \cite{Xing11}. In this sense we argue that the democratic
mixing pattern in Eq. (11) might be more natural and deserve some
more attention.

\item    One may certainly consider some possible patterns
of $U^{}_0$ which can predict a finite value of $\theta^{(0)}_{13}$
in the vicinity of the experimental value of $\theta^{}_{13}$. In
this case the three mixing angles of $U^{}_0$ may receive comparably
small corrections from the perturbation matrix $\Delta U$, and thus
the naturalness criterion can be satisfied. For example, the
following two patterns of $U^{}_0$ belong to this category
and have been discussed in the literature
\footnote{A more detailed analysis of possible forms of $U^{}_0$
has been done in Ref. \cite{RZZ}.}:
one of them is the so-called correlative mixing pattern \cite{Xing11}
\footnote{The reason for this name is simply that the three flavor
mixing angles in this constant pattern exactly satisfy the interesting
correlative relation $\theta^{(0)}_{12} + \theta^{(0)}_{13} =
\theta^{(0)}_{23}$.}
\begin{eqnarray}
U^{}_0 = \left( \matrix{
\frac{\sqrt 2}{\sqrt 3} c^{}_* & \frac{1}{\sqrt 3} c^{}_* &
s^{}_* e^{-i\delta} \cr
-\frac{1}{\sqrt 6} - \frac{1}{\sqrt 3} s^{}_* e^{i\delta} &
\frac{1}{\sqrt 3} - \frac{1}{\sqrt 6} s^{}_* e^{i\delta} &
\frac{1}{\sqrt 2} c^{}_* \cr
\frac{1}{\sqrt 6} - \frac{1}{\sqrt 3} s^{}_* e^{i\delta} &
-\frac{1}{\sqrt 3} - \frac{1}{\sqrt 6} s^{}_* e^{i\delta} &
\frac{1}{\sqrt 2} c^{}_* \cr} \right) \;
\end{eqnarray}
with $c^{}_* \equiv \cos\theta^{}_* = (\sqrt{2} +1)/\sqrt{6}$ and
$s^{}_* \equiv \sin\theta^{}_* = (\sqrt{2} -1)/\sqrt{6}$, which
predicts $\theta^{(0)}_{12} = \arctan(1/\sqrt{2}) \simeq
35.3^\circ$, $\theta^{(0)}_{23} = 45^\circ$ and $\theta^{(0)}_{13} =
\theta^{(0)}_{23} - \theta^{(0)}_{12} \simeq 9.7^\circ$; and the
other is the tetra-maximal mixing pattern \cite{Xing08}
\begin{eqnarray}
U^{}_0 = \left( \matrix{ \frac{2 + \sqrt 2}{4} &
\frac{1}{2} & \frac{2 - \sqrt 2}{4} \cr
-\frac{\sqrt{2}}{4} + \frac{i \left(\sqrt{2} - 1\right)}{4} &
\frac{1}{2} - \frac{i\sqrt{2}}{4} &
\frac{\sqrt{2}}{4} + \frac{i \left(\sqrt{2} + 1\right)}{4} \cr
-\frac{\sqrt{2}}{4} - \frac{i \left(\sqrt{2} - 1\right)}{4} &
\frac{1}{2} + \frac{i\sqrt{2}}{4} & \frac{\sqrt{2}}{4} -
\frac{i \left(\sqrt{2} + 1 \right)}{4} \cr} \right) \;
\end{eqnarray}
which predicts $\theta^{(0)}_{12} = \arctan(2-\sqrt{2})\simeq
30.4^\circ$, $\theta^{(0)}_{23} = 45^\circ$ and $\theta^{(0)}_{13} =
\arcsin[(2-\sqrt{2})/4] \simeq 8.4^\circ$. Of course, whether such
constant mixing patterns can easily be derived from some underlying
flavor symmetries remains an open question.
\end{itemize}
In short, today's model building has to take the challenge
caused by the reasonably large value of $\theta^{}_{13}$ as observed
in the Daya Bay experiment \cite{DYB}.

Furthermore, it is worth mentioning that the renormalization-group
running effects or finite quantum corrections are almost impossible
to generate $\theta^{}_{13} \simeq 9^\circ$ from $\theta^{(0)}_{13}
= 0^\circ$, unless the seesaw threshold effects or other extreme
conditions are taken into account \cite{RGE}. One may therefore
consider a pattern of $U^{}_0$ with nonzero
$\theta^{(0)}_{13}$, such as the tetra-maximal mixing pattern
\cite{ZZ} or the correlative mixing pattern \cite{LX}, as a starting
point of view to calculate the radiative corrections before confronting it
with current experimental data. We shall elaborate on this idea and
examine its impact on leptonic CP violation elsewhere \cite{LX}.

\section{The minimal perturbation to $U^{}_0$}

Note that the perturbation matrix $\Delta U$ in Eq. (4) is in
general a sum of all possible perturbations to the constant flavor
mixing matrix $U^{}_0$. From the point of view of model building, it
is helpful to single out a viable $\Delta U$ whose form is as simple
as possible. To do so, let us reexpress Eq. (4) in the following
manner:
\begin{eqnarray}
U = \left(U^{}_0 + \Delta U\right) P^{}_\nu
= U^{}_0 \left({\bf 1} + \Delta U^\prime \right) P^{}_\nu =
\left({\bf 1} + \Delta U^\prime_{\rm L} \right) U^{}_0 \left({\bf 1}
+ \Delta U^\prime_{\rm R} \right) P^{}_\nu  \; ,
\end{eqnarray}
where $\Delta U = U^{}_0 \Delta U^\prime = \Delta U^\prime_{\rm L}
U^{}_0 + U^{}_0 \Delta U^\prime_{\rm R} + \Delta U^\prime_{\rm L}
U^{}_0 \Delta U^\prime_{\rm R}$ holds, and it satisfies the
condition $U^{}_0 \Delta U^\dagger + \Delta U U^\dagger_0 + \Delta U
\Delta U^\dagger = {\bf 0}$ as a result of the unitarity of $U$
itself. Therefore, one may achieve a viable but minimal perturbation
to $U^{}_0$ by switching off $\Delta U^\prime_{\rm L}$ (or $\Delta
U^\prime_{\rm R}$) and adjusting $\Delta U^\prime_{\rm R}$ (or
$\Delta U^\prime_{\rm L}$) to its simplest form which is allowed by
current experimental data. Such a treatment is actually equivalent
to multiply $U^{}_0$ by a unitary perturbation matrix, which may
more or less deviate from the identity matrix $\bf 1$, from either
its left-hand side or its right-hand side. The first example of this
kind was given in Ref. \cite{FX96} for the democratic mixing
pattern, and its $\Delta U$ was mainly responsible for the
generation of nonzero $\theta^{}_{13}$ and $\delta$.

Here we concentrate on the typical patterns of $U^{}_0$ discussed
above and outline the main ideas of choosing the minimal
perturbations to them.
\begin{itemize}
\item     If $U^{}_0$ predicts $\theta^{(0)}_{23} =
45^\circ$ and $\theta^{(0)}_{13} =0^\circ$ together with $\theta^{(0)}_{12}
>34^\circ$ (the best-fit value based on current neutrino oscillation
data \cite{FIT}), then the simplest way to generate a relatively
large $\theta^{}_{13}$, keep $\theta^{}_{23} = \theta^{(0)}_{23} =
45^\circ$ unchanged and correct $\theta^{(0)}_{12}$ to a slightly
smaller value is to choose a complex $(2,3)$ rotation matrix as the
perturbation matrix:
\begin{eqnarray}
{\bf 1} + \Delta U^\prime = \left( \matrix{ 1 & 0 & 0 \cr 0 &
\cos\theta & i \sin\theta \cr 0 & i\sin\theta & \cos\theta \cr}
\right) ~~ {\rm or} ~~ \Delta U^\prime \simeq \left( \matrix{ 0 & 0
& 0 \cr 0 & -\frac{1}{2}\sin^2\theta & i\sin\theta \cr 0 &
i\sin\theta & -\frac{1}{2}\sin^2\theta \cr} \right) \; ,
\end{eqnarray}
where $\theta$ is a small angle to trigger the perturbation effect.
The most striking example in this category is to take $U^{}_0$ to be the
tri-bimaximal mixing pattern given in Eq. (15). The result is
\cite{Zhou07}:
\begin{eqnarray}
U = \left ( \matrix{\frac{\sqrt 2}{\sqrt 3} & \frac{1}{\sqrt 3}
\cos\theta & \frac{i}{\sqrt 3} \sin\theta \cr -\frac{1}{\sqrt 6} &
\frac{1}{\sqrt 3} \cos\theta + \frac{i}{\sqrt 2} \sin\theta &
\frac{1}{\sqrt 2} \cos\theta + \frac{i}{\sqrt 3} \sin\theta \cr
\frac{1}{\sqrt 6} & -\frac{1}{\sqrt 3} \cos\theta + \frac{i}{\sqrt
2} \sin\theta & \frac{1}{\sqrt 2} \cos\theta - \frac{i}{\sqrt 3}
\sin\theta \cr} \right ) P^{}_\nu \; ,
\end{eqnarray}
which predicts
\begin{eqnarray}
\sin^2 \theta^{}_{12} = \frac{1}{3} \left(1 - 2 \tan^2
\theta^{}_{13} \right) \; , ~~~ \sin^2 \theta^{}_{13} = \frac{1}{3}
\sin^2 \theta \; , ~~~ \theta^{}_{23} = 45^\circ \; , ~~~ \delta =
90^\circ \;
\end{eqnarray}
in the standard parametrization. Note that the obtained correlation
between $\theta^{}_{12}$ and $\theta^{}_{13}$ is especially
interesting because it leads us to $\theta^{}_{12} \to 34^\circ$
when $\theta^{}_{13} \to 9^\circ$, consistent with the present
experimental data. If $\theta^{}_{23}$ is allowed to slightly
deviate from $\theta^{(0)}_{23} =45^\circ$, then one may simply make
the replacement $i \to e^{i\delta}$ in Eq. (25).

\item     If $U^{}_0$ predicts $\theta^{(0)}_{23} =
45^\circ$ and $\theta^{(0)}_{13} =0^\circ$ together with
$\theta^{(0)}_{12} <34^\circ$, then the most economical way to
generate a relatively large $\theta^{}_{13}$, keep $\theta^{}_{23} =
\theta^{(0)}_{23} = 45^\circ$ unchanged and correct
$\theta^{(0)}_{12}$ to a slightly larger value is to choose a
complex $(1,3)$ rotation matrix as the perturbation matrix:
\begin{eqnarray}
{\bf 1} + \Delta U^\prime = \left( \matrix{ \cos\theta & 0 &
i\sin\theta \cr 0 & 1 & 0 \cr i\sin\theta & 0 & \cos\theta \cr}
\right) ~~ {\rm or} ~~ \Delta U^\prime \simeq \left( \matrix{
-\frac{1}{2}\sin^2\theta & 0 & i\sin\theta \cr 0 & 0 & 0 \cr
i\sin\theta & 0 & -\frac{1}{2}\sin^2\theta \cr} \right) \; .
\end{eqnarray}
Taking $U^{}_0$ to be the golden-ratio mixing pattern given in Eq.
(17) for example
\footnote{An interesting example with $U^{}_0$ being the
tri-bimaximal mixing pattern has been discussed in Ref.
\cite{XZZ06}, but this ansatz predicts $\theta^{}_{12}$ to be
slightly larger than $\theta^{(0)}_{12} \simeq 35.3^\circ$.},
we immediately arrive at
\begin{eqnarray}
U = \left( \matrix{ \frac{\sqrt 2}{\sqrt{5 - \sqrt 5}} \cos\theta &
\frac{\sqrt 2}{\sqrt{5 + \sqrt 5}}  & \frac{i\sqrt 2}{\sqrt{5 -
\sqrt 5}} \sin\theta \cr -\frac{1}{\sqrt{5 + \sqrt 5}} \cos\theta +
\frac{i}{\sqrt 2} \sin\theta & \frac{1}{\sqrt{5 - \sqrt 5}} &
\frac{1}{\sqrt 2} \cos\theta - \frac{i}{\sqrt{5 + \sqrt 5}}
\sin\theta \cr \frac{1}{\sqrt{5 + \sqrt 5}} \cos\theta +
\frac{i}{\sqrt 2} \sin\theta & -\frac{1}{\sqrt{5 - \sqrt 5}} &
\frac{1}{\sqrt 2} \cos\theta + \frac{i}{\sqrt{5 + \sqrt 5}}
\sin\theta \cr} \right) P^{}_\nu \; ,
\end{eqnarray}
whose predictions include $\theta^{}_{23} = 45^\circ$,
$\delta = 90^\circ$, and
\begin{eqnarray}
\sin^2\theta^{}_{12} = \frac{2}{5 + \sqrt 5} \left( 1 +
\tan^2\theta^{}_{13} \right) \; , ~~~~~
\sin^2\theta^{}_{13} = \frac{2}{5 - \sqrt 5} \sin^2\theta
\end{eqnarray}
in the standard parametrization of $U$. In this case the correlation
between $\theta^{}_{12}$ and $\theta^{}_{13}$ leads us to
$\theta^{}_{12} \to 32^\circ$ when $\theta^{}_{13} \to 9^\circ$,
compatible with current experimental data. Again, the replacement $i
\to e^{i\delta}$ in Eq. (28) allows one to obtain a somewhat more
flexible value of $\theta^{}_{23}$ which may slightly deviate from
$\theta^{(0)}_{23} =45^\circ$.

\item     If $U^{}_0$ is quite far away from the
realistic MNSP matrix $U$, one has to consider a somewhat
complicated perturbation matrix including two rotation angles. In
the neglect of CP violation, for instance, we may consider
\begin{eqnarray}
{\bf 1} + \Delta U^\prime = \left( \matrix{ c^\prime_{12} &
-s^\prime_{12} & 0 \cr s^\prime_{12} c^\prime_{23} & c^\prime_{12}
c^\prime_{23} & s^\prime_{23} \cr s^\prime_{12} s^\prime_{23} &
c^\prime_{12} s^\prime_{23} & -c^\prime_{23} \cr} \right) \; ,
\end{eqnarray}
where $c^\prime_{ij} \equiv \cos\theta^\prime_{ij}$ and
$s^\prime_{ij} \equiv \sin\theta^\prime_{ij}$ (for $ij = 12, 23$).
However, we hope that the resulting structure of $U$ still allows us
to obtain one or two predictions, in particular for the mixing
angle $\theta^{}_{13}$. An example of this kind has been given in
Ref. \cite{Xing12} by taking $U^{}_0$ to be the democratic mixing
pattern, and it predicts an interesting relationship
between $\theta^{}_{13}$ and $\theta^{}_{23}$ in the standard
parametrization:
\begin{eqnarray}
\sin\theta^{}_{13} = \frac{\sqrt{2} - \tan\theta^{}_{23}}
{\sqrt{5 - 2\sqrt{2} \tan\theta^{}_{23} + 4 \tan^2 \theta^{}_{23}}} \; .
\end{eqnarray}
Typically taking $\theta^{}_{23} \simeq 45^\circ$, we arrive at
$\theta^{}_{13} \simeq 9.6^\circ$ \cite{Xing12}, which is in
agreement with the Daya Bay result \cite{DYB}. It is easy to
accommodate a CP-violating phase in $\Delta U^\prime$ \cite{Xing12},
although its form might not be really minimal anymore.
\end{itemize}
For those constant flavor mixing patterns with $\theta^{(0)}_{13}
\neq 0^\circ$ from the very beginning, such as the correlative
\cite{Xing11} and tetra-maximal \cite{Xing08} mixing scenarios given
in Eqs. (21) and (22), the similar minimal perturbations can be
introduced in order to make the resulting MNSP matrix $U$ fit the
experimental data to a much better degree of accuracy.

It should be noted that the above discussions about possible
patterns of $\Delta U$ (or $\Delta U^\prime$) with respect to those
of $U^{}_0$ are purely phenomenological. From the point of view of
model building, it is more meaningful to consider the textures of
lepton mass matrices
\begin{eqnarray}
M^{}_l = M^{(0)}_l + \Delta M^{}_l \; , ~~~~~~~~~~~~
M^{}_\nu = M^{(0)}_\nu + \Delta M^{}_\nu \; ,
\end{eqnarray}
where $M^{(0)}_l$ and $M^{(0)}_\nu$ can be obtained in the limit of
certain flavor symmetries, and their special structures allow us to
achieve a constant flavor mixing pattern $U^{}_0$. The perturbation
matrices $\Delta M^{}_l$ and $\Delta M^{}_\nu$ play an important
role in transforming $U^{}_0$ into the realistic MNSP matrix $U$,
and thus their textures should be determined in a simple way and
with a good reason. The connection between $\Delta M^{}_{l,\nu}$ and
$\Delta U$ (or $\Delta U^\prime$) depends on the details of a lepton
flavor model and may not be very transparent in most cases. In the
basis where $M^{}_l$ is real and positive, however, $\Delta
M^{}_\nu$ can be formally expressed as
\begin{eqnarray}
\Delta M^{}_\nu = \left(U^{}_0 + \Delta U\right) \overline{M}^{}_\nu
\left(U^{}_0 + \Delta U\right)^T - U^{}_0 \overline{M}^{(0)}_\nu
U^{T}_0 \; ,
\end{eqnarray}
in which $\overline{M}^{}_\nu = P^{}_\nu \widehat{M}^{}_\nu P^T_\nu$
and $\overline{M}^{(0)}_\nu = P^{\prime}_\nu
\widehat{M}^{\prime}_\nu P^{\prime T}_\nu$ together with
$\widehat{M}^{\prime}_\nu \equiv {\rm Diag}\{m^{\prime}_1,
m^{\prime}_2, m^{\prime}_3\}$ and $P^\prime_\nu \equiv {\rm
Diag}\{e^{i\rho^\prime}, e^{i\sigma^\prime}, 1\}$. Here
$m^{\prime}_i$ (for $i=1,2,3$) denote the eigenvalues of
$M^{(0)}_\nu$ in the symmetry limit, while $\rho^\prime$ and
$\sigma^\prime$ stand for the Majorana phases in the same limit. It
is therefore possible, at least in principle, to fix the structure
of $\Delta M^{}_\nu$ with the help of a certain flavor symmetry and
current experimental data.

\section{On $\mu$-$\tau$ symmetry and CP violation}

Let us proceed to discuss two other flavor issues in the lepton
sector after the successful measurement of the smallest mixing angle
$\theta^{}_{13}$ \cite{DYB}. One of them is about a possible
departure of the largest mixing angle $\theta^{}_{23}$ from
$45^\circ$, and the other is about the strength of leptonic CP
violation. The former is an important issue in neutrino
phenomenology, because it crucially determines the structure of the
MNSP matrix $U$; and the latter is certainly more important because
the observed matter-antimatter asymmetry of the Universe might be
associated with leptonic CP violation at low energies via the seesaw
and leptogenesis mechanisms \cite{FY}.

It is well known that $\theta^{}_{23} \simeq 45^\circ$ is favored by
current atmospheric and accelerator neutrino oscillation data
\cite{PDG10}. If $\theta^{}_{23}$ is exactly equal to $45^\circ$,
then one may arrive at a partial $\mu$-$\tau$ permutation symmetry
in the MNSP matrix $U$ (i.e., the equality $|U^{}_{\mu 3}| =
|U^{}_{\tau 3}|$). This point can easily be seen from Eq. (10),
where $\theta^{}_{23} \simeq 45^\circ$ has typically been input. The
full $\mu$-$\tau$ symmetry of $U$ in modulus is described by the
equalities
\begin{eqnarray}
|U^{}_{\mu 1}| = |U^{}_{\tau 1}| \; , ~~~~
|U^{}_{\mu 2}| = |U^{}_{\tau 2}| \; , ~~~~
|U^{}_{\mu 3}| = |U^{}_{\tau 3}| \; ,
\end{eqnarray}
equivalent to two independent sets of conditions in the standard
parametrization \cite{XZ08}:
\begin{eqnarray}
\theta^{}_{23} = 45^\circ \; , ~~~ \theta^{}_{13} = 0^\circ \; ,
\end{eqnarray}
or
\begin{eqnarray}
\theta^{}_{23} = 45^\circ \; ,
~~~ \delta = \pm 90^\circ \; .
\end{eqnarray}
One can see that the constant mixing patterns in Eqs. (13), (15),
(17) and (19) satisfy the conditions in Eq. (35), while those in
Eqs. (22), (25) and (28) satisfy the conditions in Eq. (36)
\footnote{The correlative mixing pattern in Eq. (21) may also
satisfy the conditions in Eq. (36) if its CP-violating phase
$\delta$ is taken to be $\pm 90^\circ$.}.
Hence these seven scenarios of the MNSP matrix $U$ all have the
complete $\mu$-$\tau$ symmetry in modulus, or equivalently the equalities
$|U^{}_{\mu i}| = |U^{}_{\tau i}|$ (for $i=1,2,3$). Now that
$\theta^{}_{13} \neq 0^\circ$ has firmly been established by the
Daya Bay experiment \cite{DYB}, we are therefore concerned about a
possible deviation of $\theta^{}_{23}$ from $45^\circ$ and (or) a
possible departure of $\delta$ from $\pm 90^\circ$. We speculate
that $U$ might have an approximate $\mu$-$\tau$ symmetry with
$|U^{}_{\mu i}| \simeq |U^{}_{\tau i}|$, in contrast with the
approximate off-diagonal symmetry of the CKM matrix $V$ in modulus
(i.e., $|V^{}_{us}| \simeq |V^{}_{cd}|$, $|V^{}_{cb}| \simeq
|V^{}_{ts}|$ and $|V^{}_{ub}| \simeq |V^{}_{td}|$ \cite{PDG10}).

In the basis where the flavor eigenstates of three charged leptons
are identified with their mass eigenstates (i.e., $M^{}_l =
\widehat{M}^{}_l$), the Majorana neutrino mass matrix of the form
\begin{eqnarray}
M^{}_\nu = \left( \matrix{a & b & -b \cr b &
\hspace{0.17cm} c \hspace{0.17cm} & d \cr -b & d & c \cr}
\right) \;
\end{eqnarray}
predicts the $\mu$-$\tau$ permutation symmetry of the MNSP
matrix $U$ with
$\theta^{}_{13} = 0^\circ$ and $\theta^{}_{23} = 45^\circ$;
while the mass matrix of the form
\begin{eqnarray}
M^{}_\nu = \left( \matrix{a & b & - b^* \cr b & c & d \cr
- b^* & d & c^* \cr} \right) \;
\end{eqnarray}
leads us to the the $\mu$-$\tau$ symmetry of $U$ with $\delta = \pm
90^\circ$ and $\theta^{}_{23} = 45^\circ$. In either of the above
textures of $M^{}_\nu$, its entries have certain kinds of linear
correlations or equalities and thus can be generated from some
underlying flavor symmetries. In view of the experimental evidence
for $\theta^{}_{13} \neq 0^\circ$ \cite{DYB}, the pattern of
$M^{}_\nu$ in Eq. (37) has to be modified. For a similar reason, the
more reliable and accurate experimental knowledge on
$\theta^{}_{23}$ and $\delta$ will be extremely useful for us to
identify the effect of $\mu$-$\tau$ symmetry breaking and build more
realistic models for lepton mass generation, flavor mixing and CP
violation.

The fact that $\theta^{}_{13}$ is not strongly suppressed is
certainly a good news to the experimental attempts towards a
measurement of CP violation in the lepton sector. The strength of CP
violation in neutrino oscillations is described by the Jarlskog
rephasing invariant \cite{J}
\begin{eqnarray}
{\cal J}^{}_l = {\rm Im}\left(U^{}_{e1} U^{}_{\mu 2} U^*_{e2}
U^*_{\mu 1}\right) = {\rm Im}\left(U^{}_{e2} U^{}_{\mu 3} U^*_{e3}
U^*_{\mu 2}\right) = \cdots = c^{}_{12} s^{}_{12} c^2_{13} s^{}_{13}
c^{}_{23} s^{}_{23} \sin\delta \; ,
\end{eqnarray}
which is proportional to the sine of the smallest flavor mixing angle
$\theta^{}_{13}$. In the quark sector one has determined the
corresponding Jarlskog invariant ${\cal J}^{}_q \simeq 3\times
10^{-5}$ \cite{PDG10} and attributed its smallness to the strongly
suppressed values of quark flavor mixing angles (i.e.,
$\vartheta^{}_{\rm C} \equiv \vartheta^{}_{12} \simeq 13^\circ$,
$\vartheta^{}_{13} \simeq 0.2^\circ$ and $\vartheta^{}_{23} \simeq
2.4^\circ$). In the lepton sector both $\theta^{}_{12}$ and
$\theta^{}_{23}$ are large, and thus it is possible to achieve a
relatively large value of ${\cal J}^{}_l$ if the CP-violating phase
$\delta$ is not suppressed either. Note that the maximal value of
${\cal J}^{}_l$ or ${\cal J}^{}_q$ can be obtained only when the
MNSP (or CKM) matrix takes the special Cabibbo texture $V^{}_{\rm
C}$ \cite{Cabibbo78} or its equivalent form $V^{\prime}_{\rm C}$ in
the standard-parametrization phase convention:
\begin{eqnarray}
~ V^{}_{\rm C} = \left( \matrix{ \frac{1}{\sqrt 3} & \frac{1}{\sqrt
3} & \frac{1}{\sqrt 3} \cr \frac{1}{\sqrt 3} & \frac{\omega}{\sqrt
3} & \frac{\omega^2}{\sqrt 3} \cr \frac{1}{\sqrt 3} &
\frac{\omega^2}{\sqrt 3} & \frac{\omega}{\sqrt 3} \cr} \right) ~~~~
\Longrightarrow ~~~~ V^{\prime}_{\rm C} = \left( \matrix{
\frac{1}{\sqrt 3} & \frac{1}{\sqrt 3} & \frac{-i}{\sqrt 3} \cr
-\frac{1}{2} \left(1 + \frac{i}{\sqrt 3}\right) & \frac{1}{2}
\left(1 - \frac{i}{\sqrt 3}\right) & \frac{1}{\sqrt 3} \cr
\frac{1}{2} \left(1 - \frac{i}{\sqrt 3}\right) & -\frac{1}{2}
\left(1 + \frac{i}{\sqrt 3}\right) & \frac{1}{\sqrt 3} \cr} \right)
\; ,
\end{eqnarray}
where $\omega = e^{i2\pi/3}$ is the complex cube-root of unity
(i.e., $\omega^3 =1$). Therefore, $V^{}_{\rm C}$ or $V^{\prime}_{\rm
C}$ predicts $\theta^{}_{12} = \theta^{}_{23} = 45^\circ$,
$\theta^{}_{13} = \arctan(1/\sqrt{2}) \simeq 35.3^\circ$ and $\delta
= 90^\circ$, leading to the maximal CP violation ${\cal J}^{}_{\rm
max} = 1/(6\sqrt{3}) \simeq 9.6 \times 10^{-2}$. Unfortunately, both
the CKM matrix $V$ and the MNSP matrix $U$ are remarkably different
from the Cabibbo matrix $V^{}_{\rm C}$. We see ${\cal J}^{}_q/{\cal
J}^{}_{\rm max} \simeq 3\times 10^{-4}$, and hence CP violation is
rather weak in the quark sector. Taking $\theta^{}_{12} \simeq
34^\circ$, $\theta^{}_{13} \sim 9^\circ$ and $\theta^{}_{23} \simeq
45^\circ$ as a realistic example of $U$, we arrive at ${\cal J}^{}_l/{\cal
J}^{}_{\rm max} \simeq 0.37 \sin\delta$, implying that the magnitude
of leptonic CP violation can actually reach the percent level in
neutrino oscillations if the CP-violating phase $\delta$ is not
strongly suppressed (e.g., $\delta \gtrsim 16^\circ$ for the values
of three mixing angles taken above). Whether CP violation is
significant or not turns out to be an important question in lepton
physics, especially in neutrino phenomenology.

Note that ${\cal J}^{}_l \neq 0$ is a necessary and sufficient
condition for leptonic CP violation. In particular, the determinant
of the commutator of lepton mass matrices \cite{XingZZ2001}
\begin{eqnarray}
&& {\rm Det} \left(i \left[ M^{}_\nu M^\dagger_\nu,
M^{}_l M^\dagger_l
\right] \right) \nonumber \\
& = & 2 {\cal J}^{}_l \left(m^2_e - m^2_\mu \right) \left( m^2_\mu -
m^2_\tau \right) \left( m^2_\tau - m^2_e\right) \left(m^2_1 - m^2_2
\right) \left( m^2_2 - m^2_3 \right) \left( m^2_3 - m^2_1\right) \;\;
\end{eqnarray}
is unable to provide us with any more information about CP
violation. The reason is simply that ${\cal J}^{}_l$ would
automatically vanish if the masses of two charged leptons or
two neutrinos became degenerate \cite{Mei}. In other words,
one may consider the conditions for CP violation either at
the level of lepton flavor mixing (i.e., ${\cal J}^{}_l$ or
$\delta$) or at the level of lepton mass matrices, but a
confusion or double-counting problem may occur if the conditions
obtained at two different levels are mixed like Eq. (41). The
same observation is true in the quark sector, as already pointed
out in Ref. \cite{FXNucl}.

A geometrical description of CP violation in terms of the unitarity
triangles has proved very useful in the quark sector \cite{PDG10}.
This language was first applied to the lepton sector in Ref.
\cite{FX00}, in which six leptonic unitarity triangles have been
named as $\triangle^{}_e$, $\triangle^{}_\mu$, $\triangle^{}_\tau$
and $\triangle^{}_1$, $\triangle^{}_2$, $\triangle^{}_3$ (see FIG. 1
for illustration). They totally have nine independent inner angles
and eighteen independent sides, but their areas are all equal to
${\cal J}^{}_l/2$ as dictated by the unitarity of $U$ itself
\footnote{If the unitarity of $U$ is directly violated in the
presence of light sterile neutrinos or indirectly broken due to the
existence of heavy sterile neutrinos, such unitarity
triangles will change to the quadrangles \cite{Guo}
or polygons \cite{Xing3+3} in which
new CP-violating effects must be included.}.
If $U = V^{}_{\rm C}$ is taken, then the six unitarity triangles are
congruent with one another and converge to an equilateral triangle
whose sides are all equal to $1/3$ and whose area is equal to ${\cal
J}^{}_{\rm max}/2$. The fact that $U$ is rather different from
$V^{}_{\rm C}$ means somewhat smaller CP-violating effects in the
leptonic charged-current interactions. Given $\delta \simeq
90^\circ$ together with $\theta^{}_{12} \simeq 34^\circ$,
$\theta^{}_{13} \sim 9^\circ$ and $\theta^{}_{23} \simeq 45^\circ$,
for instance, the nine inner angles of the six unitarity triangles
in FIG. 1 turn out to be
\begin{eqnarray}
\Phi \equiv \left( \matrix{ \Phi^{}_{e1} & \Phi^{}_{e2} & \Phi^{}_{e3} \cr
\Phi^{}_{\mu 1} & \Phi^{}_{\mu 2} & \Phi^{}_{\mu 3} \cr
\Phi^{}_{\tau 1} & \Phi^{}_{\tau 2} & \Phi^{}_{\tau 3} \cr} \right) \simeq
\left( \matrix{12.05^\circ & 26.11^\circ & 141.8^\circ \cr
83.98^\circ & 76.94^\circ & 19.08^\circ \cr 83.98^\circ &
76.94^\circ & 19.08^\circ \cr} \right) \; .
\end{eqnarray}
We see that this unitarity-triangle angle matrix exhibits an
interesting $\mu$-$\tau$ symmetry as guaranteed by the inputs
$\delta \simeq 90^\circ$ and $\theta^{}_{23} \simeq 45^\circ$. In
addition, its nine matrix elements are rephasing-invariant and
satisfy the sum rules \cite{Luo12}
\begin{eqnarray}
\sum_\alpha \Phi^{}_{\alpha i} = \sum_i \Phi^{}_{\alpha i} = 180^\circ \; ,
\end{eqnarray}
where the subscript $\alpha$ runs over $e$, $\mu$ and $\tau$, and
$i$ runs over $1$, $2$ and $3$. We expect that the future
long-baseline neutrino oscillation experiments can hopefully
determine or constrain some of the above angles and thus pin down
the CP-violating phase $\delta$ of $U$ even in the presence of
terrestrial matter effects on the unitarity triangles \cite{Xing04}.

\section{Nine distinct parametrizations}

The $3\times 3$ unitary flavor mixing matrix can always be
parametrized in terms of three rotation angles and a few phase
angles. A classification of all the possible parametrizations of
this kind has been done in Ref. \cite{FX98}. Here we list nine
topologically distinct parametrizations of the MNSP matrix $U$ in
TABLE I, in which three rotation matrices are defined as
\begin{eqnarray}
R^{}_{12}(\theta^{}_{12}, \delta) & = & \left( \matrix{ c^{}_{12} &
s^{}_{12} & 0 \cr -s^{}_{12} & c^{}_{12} & 0 \cr 0 & 0 &
e^{-i\delta} \cr}
\right) \; , \nonumber \\
R^{}_{23}(\theta^{}_{23}, \delta) & = & \left( \matrix{ e^{-i\delta}
& 0 & 0 \cr 0 & c^{}_{23} & s^{}_{23} \cr 0 & -s^{}_{23} & c^{}_{23}
\cr}
\right) \; , \nonumber \\
R^{}_{13}(\theta^{}_{13}, \delta) & = & \left( \matrix{ c^{}_{13} &
0 & s^{}_{13} \cr 0 & e^{-i\delta} & 0 \cr -s^{}_{13} & 0 &
c^{}_{13} \cr} \right) \; ,
\end{eqnarray}
with $c^{}_{ij} \equiv \cos\theta^{}_{ij}$ and $s^{}_{ij} \equiv
\sin\theta^{}_{ij}$ (for $ij = 12, 13, 23$). Although all the
parametrizations of $U$ (or the CKM matrix $V$) are mathematically
equivalent, we argue that some of them might be phenomenologically
more interesting in the sense that they might either make the
underlying dynamics of flavor mixing more transparent or lead to
more straightforward and simpler relations between fundamental
parameters and observable quantities \cite{FX98}. In other words,
they are possible to provide us with some novel points of view on
the structure of lepton or quark flavor mixing. As stressed by
Feynman, ``different views suggest different kinds of modifications
which might be made" and ``a good theoretical physicist today might
find it useful to have a wide range of physical viewpoints and
mathematical expressions of the same theory available to him"
\cite{Feynman}.

Let us focus on the MNSP matrix $U$ and make some comments on its
nine different parametrizations listed in TABLE I.
\begin{itemize}
\item     Pattern (1) was first proposed in Ref. \cite{FX97} ,
and it is usually expressed in terms of the following notations:
\begin{eqnarray}
U = \left ( \matrix{ s^{}_l s^{}_{\nu} c + c^{}_l c^{}_{\nu}
e^{-i\varphi} & s^{}_l c^{}_{\nu} c - c^{}_l s^{}_{\nu}
e^{-i\varphi} & s^{}_l s \cr c^{}_l s^{}_{\nu} c - s^{}_l c^{}_{\nu}
e^{-i\varphi} & c^{}_l c^{}_{\nu} c + s^{}_l s^{}_{\nu}
e^{-i\varphi} & c^{}_l s \cr - s^{}_{\nu} s   & - c^{}_{\nu} s   & c
\cr } \right ) P^{}_\nu \; ,
\end{eqnarray}
where $c^{}_{l,\nu} \equiv \cos\theta^{}_{l,\nu}$, $s^{~}_{l,\nu}
\equiv \sin\theta^{}_{l,\nu}$, $c \equiv \cos\theta$ and $s \equiv
\sin\theta$. In the leading-order approximation we have $s^{}_\nu
\simeq s^{}_{12}$, $s \simeq s^{}_{23}$ and $s^{}_l \simeq
s^{}_{13}/s^{}_{23}$. There are two remarkable merits of this
parametrization: 1) it is quite useful for model building if the
neutrino mass spectrum has a normal hierarchy as the charged-lepton
or quark mass spectrum (e.g., $\tan\theta^{}_l \simeq
\sqrt{m^{}_e/m^{}_\mu}$ and $\tan\theta^{}_\nu \simeq
\sqrt{m^{}_1/m^{}_2}$ have been conjectured in Ref. \cite{FX06});
and 2) it allows us to obtain impressively simple expressions of the
one-loop renormalization-group equations for three flavor mixing
angles and three CP-violating phases, much simpler than those
obtained by using the standard parametrization in Eq. (2)
\cite{Xing06}.

\item     Pattern (2) is equivalent to the original Kobayashi-Maskawa
parametrization \cite{CKM}. The structure of this pattern and those
of patterns (3) and (7) have a common feature: the rotation matrix
on the left-hand side of $U$ is $R^{}_{23}(\theta^{}_{23})$, which
is commutable with a diagonal matrix of the form ${\rm Diag}\{A, 0,
0\}$. When a neutrino beam travels through a normal medium, the
coherent forward scattering effect induced by the charged-current
interactions of electron neutrinos (or antineutrinos) with matter
can just generate such an effective potential term \cite{MSW}. Hence
patterns (2), (3) and (7) are more convenient to describe matter
effects on neutrino oscillations. In particular, it has been
shown in Ref. \cite{Zhou} that these three parametrizations of $U$
can all lead us to the exact and interesting Toshev relation
\cite{Toshev}
\begin{eqnarray}
\sin\tilde{\delta} \sin 2\tilde{\theta}^{}_{23} =
\sin\delta \sin 2\theta^{}_{23} \; ,
\end{eqnarray}
where $\tilde{\theta}^{}_{23}$ and $\tilde{\delta}$ denote the
effective counterparts of $\theta^{}_{23}$ and $\delta$ in matter.

\item     Pattern (3) is equivalent to the standard parametrization
of $U$ given in Eq. (2) \cite{PDG10}, although its phase convention
is slightly different. This representation becomes most popular
today because its three mixing angles ($\theta^{}_{12}$,
$\theta^{}_{23}$, $\theta^{}_{13}$) directly measure the effects of
solar, atmospheric and reactor neutrino oscillations ($\sin^2
2\theta^{}_{12}$, $\sin^2 2 \theta^{}_{23}$, $\sin^2 2
\theta^{}_{13}$) in the two-flavor approximation in vacuum.
Furthermore, the smallest mixing angle $\theta^{}_{13}$ determines
the smallest matrix element $U^{}_{e3}$ of the MNSP matrix $U$ in a
way analogous to the standard parametrization of the CKM matrix $V$,
where the smallest element $V^{}_{ub}$ is controlled by the
smallest mixing angle $\vartheta^{}_{13}$ \cite{PDG10}. Hence in
this parametrization the hierarchy of three mixing angles can almost
truly reflect the overall hierarchy of the flavor mixing matrix, as
we have discussed in sections III, IV and V.

\item     Pattern (5) is structurally special in the sense that only the
$3\times 3$ flavor mixing matrix $U$ can have this form around its
``central element" $U^{}_{\mu 2}$. As a result, two off-diagonal
asymmetries of $U$ in modulus can simply be expressed as
\begin{eqnarray}
{\cal A}^{}_{\rm L} & \equiv &
|U^{}_{e2}|^2 - |U^{}_{\mu 1}|^2 = |U^{}_{\mu 3}|^2 - |U^{}_{\tau
2}|^2 = |U^{}_{\tau 1}|^2 - |U^{}_{e3}|^2
= s^2_{12} \left( c^2_{13} - c^{\prime 2}_{13} \right) \; ,
\nonumber \\
{\cal A}^{}_{\rm R} & \equiv &
|U^{}_{e2}|^2 - |U^{}_{\mu 3}|^2 =
|U^{}_{\mu 1}|^2 - |U^{}_{\tau 2}|^2 =
|U^{}_{\tau 3}|^2 - |U^{}_{e1}|^2 =
s^2_{12} \left( c^2_{13} - s^{\prime 2}_{13} \right) \; .
\end{eqnarray}
Current neutrino oscillation data indicate that both ${\cal
A}^{}_{\rm L} \neq 0$ and ${\cal A}^{}_{\rm R} \neq 0$ hold at the
$3\sigma$ level, implying that the MNSP matrix $U$ is apparently
asymmetric in modulus about either its $U^{}_{e1}$-$U^{}_{\mu
2}$-$U^{}_{\tau 3}$ axis or its $U^{}_{e3}$-$U^{}_{\mu
2}$-$U^{}_{\tau 1}$ axis \cite{XingZZ02}. In contrast, the CKM
matrix $V$ is roughly symmetric in modulus about its
$V^{}_{ud}$-$V^{}_{cs}$-$V^{}_{tb}$ axis. Another unique feature of
pattern (5) is that it assures three mixing angles to be comparably
large and the (Dirac) CP-violating phase to be nearly minimal (in
particular, $\vartheta^{}_{12} \simeq 13.2^\circ$,
$\vartheta^{}_{13} \simeq 10.1^\circ$, $\vartheta^{\prime}_{13}
\simeq 10.3^\circ$ and $\delta \simeq 1.1^\circ$ for the quark
sector; and all the three mixing angles are around $45^\circ$ for
the lepton sector with a much smaller Dirac CP-violating phase
\cite{Gerard}). In this sense the approximate flavor mixing
democracy and minimal CP violation have been discussed in Ref.
\cite{Gerard} as an alternative point of view to look at the flavor
puzzles of leptons and quarks.

\item     Some interest has also been paid to patterns (4), (6) and (8)
\cite{Huang} for two simple reasons: 1) none of the three flavor
mixing angles is suppressed in each of them; and 2) the CP-violating
phase $\delta$ is strongly correlated with the mixing angles.  This
kind of strong parameter correlation might allow one to determine
$\delta$ with fewer uncertainties from an experimental point of
view, as compared with the relatively weak parameter correlation in
patterns (3), (7) and (9), where the value of $\theta^{}_{13}$ is
much smaller than those of $\theta^{}_{12}$ and $\theta^{}_{23}$.
Generally speaking, however, patterns (4), (6), (7), (8) and (9)
seem to be somewhat less interesting than patterns (1), (2), (3) and
(5) for the phenomenological studies of flavor physics.
\end{itemize}
For each of the nine parametrizations of the MNSP matrix $U$, the
explicit expression of the Jarlskog invariant of leptonic CP
violation ${\cal J}^{}_l$ has been given in TABLE I.

\section{The Wolfenstein-like expansion}

Following the conjecture that the MNSP matrix $U$ is composed of a
constant leading term $U^{}_0$ and a perturbation term $\Delta U$ as
described in Eq. (4), we have argued that the structure of $\Delta
U$ is relatively natural if $U^{}_0$ takes the democratic mixing
pattern. In particular, the numerical result of $\Delta U$ obtained
in Eq. (12) indicates that its nine matrix elements are all of
${\cal O}(0.1)$ and thus can easily be described by a common small
parameter. This observation reminds us of the well-known Wolfenstein
parametrization of the CKM matrix $V$, which was proposed soon after
the smallest element $V^{}_{ub}$ was experimentally determined
\cite{W}. Such a parametrization has proved to be very useful
because it clearly reveals the observed strong hierarchy in the
quark flavor structure. Although a straightforward Wolfenstein-like
parametrization of the MNSP matrix $U$ has been discussed
\cite{Xing03}, it is not useful because the structure of $U$ is not as
hierarchical as that of $V$. A different starting point of view is
to speculate that the realistic form of $U$ comes from the
democratic mixing pattern $U^{}_0$ and a Wolfenstein-like
perturbation $\Delta U$. Here we proceed to explore this noteworthy
possibility in some detail, so as to illustrate an alternative way
for describing the phenomenon of lepton flavor mixing other than
those parametrizations discussed in section VI.

Comparing Eq. (11) with Eq. (2), we can define three
Wolfenstein-like parameters in the following way:
\begin{eqnarray}
\theta^{}_{12} & \equiv & \theta^{(0)}_{12} - \theta^{}_x \;\;
{\rm with} \;\; \sin\theta^{}_x \equiv \lambda \; ,
\nonumber \\
\theta^{}_{23} & \equiv & \theta^{(0)}_{23} - \theta^{}_y \;\;
{\rm with} \;\; \sin\theta^{}_y \equiv A \lambda \; ,
\nonumber \\
\theta^{}_{13} & \equiv & \theta^{(0)}_{13} - \theta^{}_z \;\;
{\rm with} \;\; \sin\theta^{}_z \equiv -B \lambda \; ,
\end{eqnarray}
where the magnitudes of $A$ and $B$ are expected to be of ${\cal
O}(1)$. In view of $\theta^{(0)}_{12} = 45^\circ$,
$\theta^{(0)}_{23} \simeq 54.7^\circ$ and $\theta^{(0)}_{13} =
0^\circ$ given by $U^{}_0$ together with $\theta^{}_{12} \simeq
34^\circ$, $\theta^{}_{23} \simeq 45^\circ$ and $\theta^{}_{13}
\simeq 9^\circ$ extracted from current neutrino oscillation data,
for example, we typically obtain
\begin{eqnarray}
\lambda \simeq 0.19 \; , \;\;\;\; A \simeq 0.88 \; , \;\;\;\;
B \simeq 0.82 \; .
\end{eqnarray}
Up to the accuracy of ${\cal O}(\lambda^2)$, the sine and cosine of
each flavor mixing angle are found to be
\begin{eqnarray}
s^{}_{12} & \simeq & \frac{1}{\sqrt 2} \left( 1 - \lambda
- \frac{1}{2} \lambda^2 \right) \; , \nonumber \\
c^{}_{12} & \simeq & \frac{1}{\sqrt 2} \left( 1 + \lambda
- \frac{1}{2} \lambda^2 \right) \; , \nonumber \\
s^{}_{13} & = & B \lambda \; , \nonumber \\
c^{}_{13} & \simeq & 1 - \frac{1}{2} B^2 \lambda^2 \; , \nonumber \\
s^{}_{23} & \simeq & \frac{\sqrt 2}{\sqrt 3} \left( 1 -
\frac{1}{\sqrt 2} A \lambda
- \frac{1}{2} A^2 \lambda^2 \right) \; , \nonumber \\
c^{}_{23} & \simeq & \frac{1}{\sqrt 3} \left( 1 + \sqrt{2} A \lambda
- \frac{1}{2} A^2 \lambda^2 \right) \; .
\end{eqnarray}
Then the nine matrix elements of $U$ can be expanded in terms of the
small parameter $\lambda$ as
\begin{eqnarray}
U^{}_{e1} & \simeq & \frac{1}{\sqrt 2} \left[ 1 + \lambda -
\frac{1}{2} \left(1 +B^2\right) \lambda^2 \right] e^{i\rho} \; ,
\nonumber \\
U^{}_{e2} & \simeq & \frac{1}{\sqrt 2} \left[ 1 - \lambda -
\frac{1}{2} \left(1 +B^2\right) \lambda^2 \right] e^{i\sigma} \; ,
\nonumber \\
U^{}_{e3} & = & \hat{B}^* \lambda \; ,
\nonumber \\
U^{}_{\mu 1} & \simeq & -\frac{1}{\sqrt 6} \left\{ 1 +
\left(\sqrt{2} A -1 + \sqrt{2}\hat{B} \right) \lambda - \frac{1}{2}
\left[1 + 2\sqrt{2} A + A^2 -2 \left(\sqrt{2} - A \right) \hat{B}
\right] \lambda^2 \right\} e^{i\rho} \; ,
\nonumber \\
U^{}_{\mu 2} & \simeq & \frac{1}{\sqrt 6} \left\{ 1 + \left(\sqrt{2}
A + 1 - \sqrt{2}\hat{B} \right) \lambda - \frac{1}{2} \left[1 -
2\sqrt{2} A + A^2 -2 \left(\sqrt{2} + A \right) \hat{B} \right]
\lambda^2 \right\} e^{i\sigma} \; ,
\nonumber \\
U^{}_{\mu 3} & \simeq & \frac{\sqrt 2}{\sqrt 3} \left[ 1 -
\frac{1}{\sqrt 2} A \lambda - \frac{1}{2} \left( A^2 +
B^2 \right) \lambda^2 \right] \; ,
\nonumber \\
U^{}_{\tau 1} & \simeq & \frac{1}{\sqrt 3} \left\{ 1 -
\frac{1}{\sqrt 2} \left(\sqrt{2} + A + \hat{B} \right) \lambda -
\frac{1}{2} \left[1 - \sqrt{2} A + A^2 + \sqrt{2} \left(1 + \sqrt{2}
A \right) \hat{B} \right] \lambda^2 \right\} e^{i\rho} \; ,
\nonumber \\
U^{}_{\tau 2} & \simeq & -\frac{1}{\sqrt 3} \left\{ 1
+\frac{1}{\sqrt 2} \left(\sqrt{2} - A + \hat{B} \right) \lambda -
\frac{1}{2} \left[1 - \sqrt{2} A + A^2 + \sqrt{2} \left(1 - \sqrt{2}
A \right) \hat{B} \right] \lambda^2 \right\} e^{i\sigma} \; ,
\nonumber \\
U^{}_{\tau 3} & \simeq & \frac{1}{\sqrt 3} \left[ 1 + \sqrt{2} A
\lambda - \frac{1}{2} \left( A^2 + B^2 \right) \lambda^2 \right] \;
,
\end{eqnarray}
where $\hat{B} \equiv B e^{i\delta}$ is defined, and the Majorana
CP-violating phases $\rho$ and $\sigma$ are included. In this
parametrization of $U$, the Jarlskog invariant of CP violation and
two off-diagonal asymmetries defined in Eq. (47) turn out to be
\begin{eqnarray}
{\cal J}^{}_l \simeq \frac{1}{3\sqrt 2} B\lambda\sin\delta
\left[ 1 + \frac{1}{\sqrt 2} A\lambda - \left( 2A^2 + B^2 -2 \right)
\lambda^2 \right] \; ,
\end{eqnarray}
and
\begin{eqnarray}
{\cal A}^{}_{\rm L} & \simeq & \frac{1}{3} \left[ 1 - \sqrt{2}
\left( \sqrt{2} + A + B\cos\delta \right) \lambda - \frac{1}{2}
\left( A^2 + 5B^2 - 4\sqrt{2} A + 2 AB\cos\delta \right) \lambda^2
\right] \; ,
\nonumber \\
{\cal A}^{}_{\rm R} & \simeq & -\frac{1}{6} \left[ 1 - 2
\left( 2\sqrt{2} A - 3\right) \lambda - \left( 2 A^2 + B^2\right)
\lambda^2 \right] \; .
\end{eqnarray}
Taking $\lambda \simeq 0.19$ and $B \simeq 0.82$ for example, we
obtain ${\cal J}^{}_l/{\cal J}^{}_{\rm max} \simeq \sqrt{6} B\lambda
\sin\delta \sim 0.38\sin\delta$ in the leading-order approximation,
consistent with our estimate made below Eq. (40). So the leptonic
Jarlskog invariant can be as large as a few percent for an
unsuppressed value of $\delta$.

\section{Summary and concluding remarks}

Motivated by the robust Daya Bay result for a relatively large value
of the smallest neutrino mixing angle $\theta^{}_{13}$, we have
explored the leptonic flavor mixing structure and CP violation in a
quite systematic way. Our main points and results are summarized as
follows.

(1) We have outlined two phenomenological strategies for
understanding the textures of lepton mass matrices and thus the
structure of lepton flavor mixing:
\begin{itemize}
\item     The MNSP matrix $U$ is expressed as the sum of
a constant leading term $U^{}_0$ and a small perturbation term
$\Delta U$. $U^{}_0$ is responsible for two larger mixing
angles and may result from a certain flavor symmetry,
while $\Delta U$ is responsible for the smallest mixing angle
and CP-violating phase(s) and can be generated from the
symmetry breaking or quantum corrections. As a consequence
of the flavor symmetry at the level of lepton mass matrices,
their entries have certain kinds of linear correlations or
equalities.

\item     The mixing angles of $U$ are associated with the ratios
of charged-lepton and neutrino masses. In this case the lepton mass
matrices may have some texture zeros which can also be derived from
a certain flavor symmetry.
\end{itemize}
At present the first strategy is more popular for model building,
but one has to come up with some new ideas in order to account for
the observed value of $\theta^{}_{13}$. We stress that both
approaches deserve further studies, in particular when the
neutrino oscillation data on three flavor mixing angles become more
and more precise.

(2) We have reexamined the democratic, bimaximal, tri-bimaximal,
golden-ratio and hexagonal mixing patterns as possible candidates
for $U^{}_0$, and constrained their respective perturbations by
using current experimental data. To generate $\theta^{}_{13} \simeq
9^\circ$ together with the allowed values of $\theta^{}_{12}$ and
$\theta^{}_{23}$, we find that the structure of $\Delta U$ with
respect to the democratic mixing pattern seems to be most natural
because its nine elements are all of ${\cal O}(0.1)$. So we have
proposed a Wolfenstein-like expansion of the MNSP matrix $U$ with
the help of the democratic mixing pattern and a small parameter
$\lambda \simeq 0.19$, as compared with the well-known Wolfenstein
parametrization of the CKM matrix $V$.

(3) Concentrating on the general conjecture $U = \left(U^{}_0 +
\Delta U\right) P^{}_\nu$, we have discussed the possibly minimal
form of $\Delta U$ for a given pattern of $U^{}_0$ as mentioned
above. The possibility of $(U^{}_0)^{}_{e 3} \neq 0$ has also been
taken into account. Let us emphasize two points in the following:
\begin{itemize}
\item     Given $(U^{}_0)^{}_{e 3} = 0$ (e.g., the
tri-bimaximal mixing pattern), the $\Delta U$ part has to be taken
more seriously than before in building a realistic model of lepton
mass matrices. The reason is simply that it is a highly nontrivial
job to generate $\theta^{}_{13} \simeq 9^\circ$ from
$\theta^{(0)}_{13} = 0^\circ$.

\item     It is worth paying more attention to the patterns
of $U^{}_0$ with nonzero $\theta^{(0)}_{13}$,
such as the correlative or tetra-maximal mixing
pattern. In this case one might be able to adjust the structure
of $\Delta U$ to a simple form, but whether the origin of
$U^{}_0$ itself has a good reason (e.g., a simple or
convincing flavor symmetry) remains an open question.
\end{itemize}
For a detailed analysis of the renormalization-group running effects
on $U$ with the value of $\theta^{}_{13}$ as observed in the Daya
Bay experiment, we refer the reader to Ref. \cite{LX}.

(4) We have pointed out a salient feature of the MNSP matrix $U$: it
may exhibit an approximate $\mu$-$\tau$ permutation symmetry
in modulus thanks
to $\theta^{}_{23} \simeq 45^\circ$. It is therefore crucial for the
future neutrino oscillation experiments to determine the departure
of $\theta^{}_{23}$ from $45^\circ$. From the point of view of model
building, the sign of $\theta^{}_{23} - 45^\circ$ is a useful
and sensitive model discriminator as the size of $\theta^{}_{13}$ is.

(5) We have stressed that $\delta \simeq \pm 90^\circ$ is not only
important for enhancing the strength of leptonic CP violation but
also helpful for making the structure of $U$ closer to its
$\mu$-$\tau$ symmetry limit. A geometrical description of CP
violation has also been highlighted by considering the language of
the leptonic unitarity triangles.

(6) We have summarized the main merits of nine topologically
distinct parametrizations of $U$. Some of them turn out to be useful
in revealing the features of lepton flavor mixing and CP violation.
We have also introduced an alternative way to describe the MNSP
matrix $U$ --- it is a Wolfenstein-like expansion of $U$ based on the
democratic mixing pattern.

Let us reiterate that the relative sizes of the nine elements
of the MNSP matrix $U$ cannot be completely fixed unless we
have known $\theta^{}_{23} > 45^\circ$ or
$\theta^{}_{23} < 45^\circ$ as well as the range of $\delta$.
With the help of the available experimental data and the
unitarity of $U$, we find
\begin{eqnarray}
|U^{}_{e 1}| > |U^{}_{\mu 3}| \sim |U^{}_{\tau 3}| >
|U^{}_{\mu 2}| \sim |U^{}_{\tau 2}| > |U^{}_{e 2}| > |U^{}_{\mu 1}|
\sim |U^{}_{\tau 1}| > |U^{}_{e 3}| \; ,
\end{eqnarray}
where ``$\sim$" implies that the relative magnitudes of
$|U^{}_{\mu i}|$ and $|U^{}_{\tau i}|$ (for $i=1,2,3$)
remain undetermined at present. In comparison, the nine elements
of the CKM matrix $V$ are known to have the following hierarchy
\cite{Xing95}:
\begin{eqnarray}
|V^{}_{tb}| > |V^{}_{ud}| > |V^{}_{cs}| \gg
|V^{}_{us}| > |V^{}_{cd}| \gg |V^{}_{cb}| > |V^{}_{ts}|
\gg |V^{}_{td}| > |V^{}_{ub}| \; .
\end{eqnarray}
We see that there is a striking similarity between the quark and lepton
flavor mixing matrices: the smallest elements of both $V$ and $U$
appear in their respective top-right corners.

It is certainly impossible to make an exhaustive overview of all the
problems associated with the leptonic flavor mixing structure and CP
violation at this stage and in this paper
\footnote{In particular, the impact of
$\theta^{}_{13} \simeq 9^\circ$ on the renormalization-group running
behaviors of three flavor mixing angles and three CP-violating
phases is not covered, nor is the issue for going beyond the
$3\times 3$ lepton flavor mixing matrix in the presence of
three light or heavy sterile neutrinos. As for these two topics, we
refer the reader to Ref. \cite{LX} and Ref. \cite{Xing3+3}
respectively.}.
But we hope that some of our points or
questions may trigger some new ideas and further efforts towards
deeper understanding of the underlying dynamics responsible for
lepton mass generation, flavor mixing and CP violation. We emphasize
that the lessons learnt from the quark sector are especially
beneficial to our attempts in the lepton sector. Let us illustrate
why this emphasis makes sense from a historical point of view as
below.

In the history of flavor physics it took quite a long time to
measure the four independent parameters of the CKM matrix $V$,
but the experimental development had a clear roadmap:
\begin{eqnarray}
\vartheta^{}_{12} ~ ({\rm or} ~ |V^{}_{us}|) ~~~ \Longrightarrow ~~~
\vartheta^{}_{23} ~ ({\rm or} ~ |V^{}_{cb}|) ~~~ \Longrightarrow ~~~
\vartheta^{}_{13} ~ ({\rm or} ~ |V^{}_{ub}|) ~~~ \Longrightarrow ~~~
\delta ~({\rm quark}) \; .
\end{eqnarray}
Namely, the observation of the largest mixing angle
$\vartheta^{}_{12}$ was the first step, the determination of the
smallest mixing angle $\vartheta^{}_{13}$ (or equivalently, the
smallest matrix element $|V^{}_{ub}|$) was an important turning
point, and then the quark flavor physics entered an era of precision
measurements in which CP violation could be explored and new physics
could be searched for. Interestingly and hopefully, the lepton
flavor physics is repeating the same story:
\begin{eqnarray}
\theta^{}_{23} ~ ({\rm or} ~ |U^{}_{\mu 3}|) ~~~ \Longrightarrow ~~~
\theta^{}_{12} ~ ({\rm or} ~ |U^{}_{e 2}|) ~~~ \Longrightarrow ~~~
\theta^{}_{13} ~ ({\rm or} ~ |U^{}_{e 3}|) ~~~ \Longrightarrow ~~~
\delta ~({\rm lepton}) \; ,
\end{eqnarray}
where $\theta^{}_{23}$ is the largest and $\theta^{}_{13}$ is the
smallest. The observation of $\theta^{}_{13}$ (or equivalently, the
smallest matrix element $|U^{}_{e3}|$) in the Daya Bay experiment is
paving the way for future experiments to study leptonic CP violation
and to look for possible
new physics (e.g., whether the $3\times 3$ MNSP matrix $U$ is
exactly unitary or not), in particular through the measurements of
neutrino oscillations for different sources of neutrino beams.
The Majorana nature of three massive
neutrinos and their other two CP-violating phases (i.e., $\rho$ and
$\sigma$) can also be probed in the new era of neutrino physics.

\acknowledgments

I would like to thank S. Luo for her technical helps, and to Y.F. Li and
S. Luo for some useful discussions. I am also indebted to J. Cao
and Y.F. Wang for many interesting ``weak" interactions.
This work was supported in part by the National Natural Science
Foundation of China under grant No. 11135009.

\begin{figure*}
\vspace{4cm}
\centering
\includegraphics[bb = 230 640 380 714,scale=0.9]{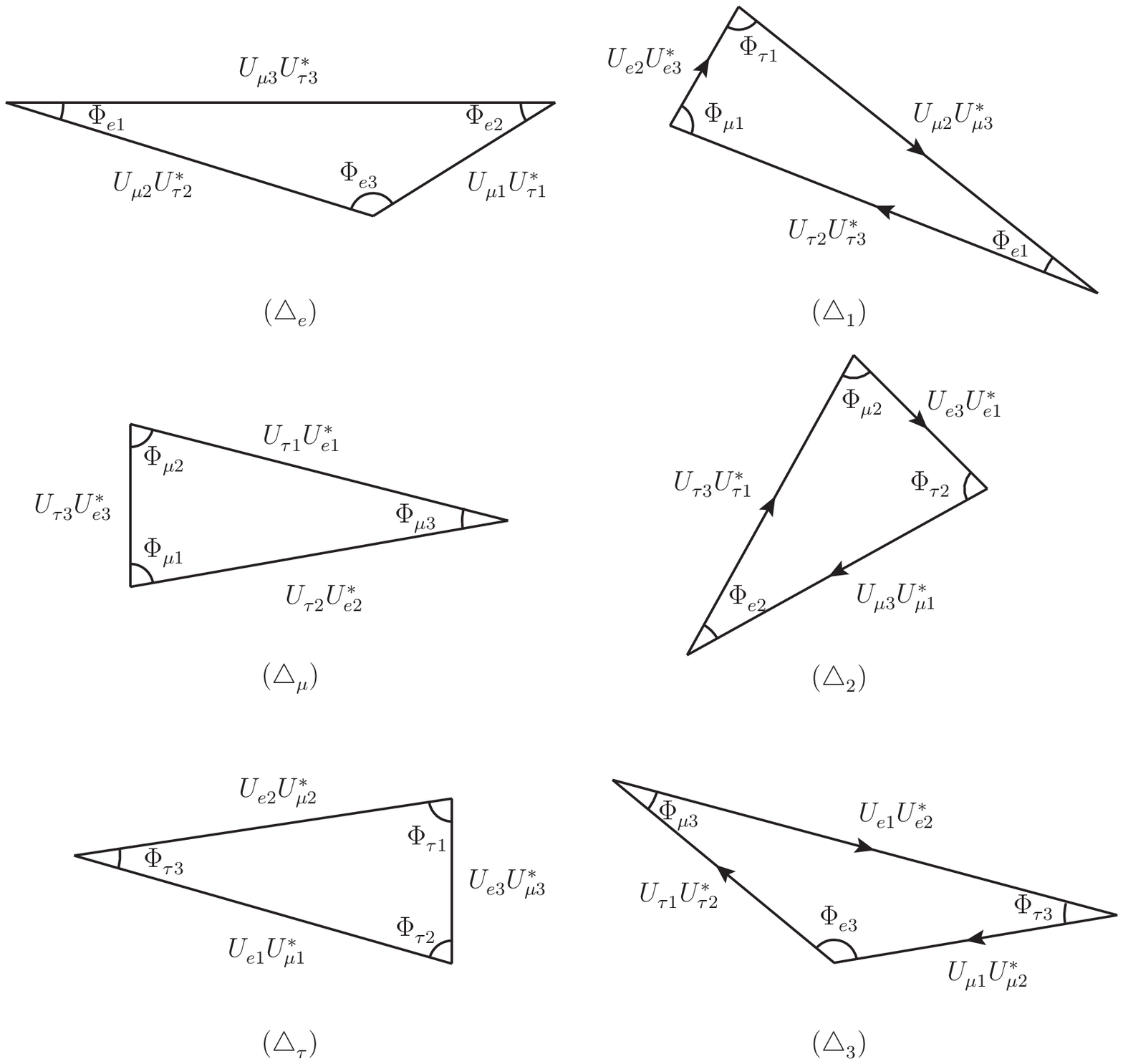}
\vspace{12cm} \caption{Six unitarity triangles of the MNSP matrix
$U$ in the complex plane. Each triangle is named by the index that
does not appear in its three sides [16], and the relative scale of
the six triangles is roughly consistent with the assumption of
$\delta \simeq 90^\circ$ and current experimental data on the three
flavor mixing angles of $U$ [45].}
\end{figure*}

\begin{table}
\caption{A classification of nine topologically distinct
parametrizations of the MNSP matrix $U$ in terms of three rotation
angles and three phase angles [47], where $P^{}_\nu = {\rm
Diag}\{e^{i\rho}, e^{i\sigma}, 1\}$ denotes the Majorana phase
matrix. The phase (or sign) convention of each parametrization is
adjustable.}
\begin{center}
\begin{tabular}{lcl}
Different parametrizations     & ~~~~ & Useful relations
\\ \hline
{Pattern (1):} ~ $U = R^{}_{12}(\theta^{}_{12}) \otimes
R^{}_{23}(\theta^{}_{23}, \delta) \otimes
R^{T}_{12}(\theta^{\prime}_{12}) \otimes P^{}_\nu$ && ${\cal
J}^{}_l = s^{~}_{12} c^{~}_{12} s^{\prime}_{12}
c^{\prime}_{12} s^2_{23} c^{}_{23} \sin\delta$ \\
$\left ( \matrix{ s^{~}_{12} s^{\prime}_{12} c^{}_{23} + c^{~}_{12}
c^{\prime}_{12} e^{-{\rm i}\delta}   & s^{~}_{12} c^{\prime}_{12}
c^{}_{23} - c^{~}_{12} s^{\prime}_{12} e^{-{\rm i}\delta}  &
s^{~}_{12} s^{}_{23} \cr c^{~}_{12} s^{\prime}_{12} c^{}_{23} -
s^{~}_{12} c^{\prime}_{12} e^{-{\rm i}\delta}  & c^{~}_{12}
c^{\prime}_{12} c^{}_{23} + s^{~}_{12} s^{\prime}_{12} e^{-{\rm
i}\delta}  & c^{~}_{12} s^{}_{23} \cr - s^{\prime}_{12} s^{}_{23}
& - c^{\prime}_{12} s^{}_{23}   & c^{}_{23} \cr} \right ) P^{}_\nu $
&& $\begin{array}{l}
\tan\theta^{}_{12} = |U^{}_{e3}/U^{}_{\mu 3}| \\
\tan\theta^{\prime}_{12} = |U^{}_{\tau 1}/U^{}_{\tau 2}| \\
\cos\theta^{}_{23} = |U^{}_{\tau 3}| \end{array} $ \\ \hline
{Pattern (2):} ~ $U = R^{}_{23}(\theta^{}_{23}) \otimes
R^{}_{12}(\theta^{}_{12}, \delta) \otimes
R^{T}_{23}(\theta^{\prime}_{23}) \otimes P^{}_\nu$ && ${\cal
J}^{}_l = s^2_{12} c^{~}_{12} s^{}_{23} c^{}_{23}
s^{\prime}_{23} c^{\prime}_{23} \sin\delta$ \\
$\left ( \matrix{ c^{~}_{12}  & s^{~}_{12} c^{\prime}_{23}    &
-s^{~}_{12} s^{\prime}_{23} \cr -s^{~}_{12} c^{}_{23}   & c^{~}_{12}
c^{}_{23} c^{\prime}_{23} + s^{}_{23} s^{\prime}_{23} e^{-{\rm
i}\delta} & -c^{~}_{12} c^{}_{23} s^{\prime}_{23} + s^{}_{23}
c^{\prime}_{23} e^{-{\rm i}\delta} \cr s^{~}_{12} s^{}_{23}    &
-c^{~}_{12} s^{}_{23} c^{\prime}_{23} + c^{}_{23} s^{\prime}_{23}
e^{-{\rm i}\delta} & c^{~}_{12} s^{}_{23} s^{\prime}_{23} +
c^{}_{23} c^{\prime}_{23} e^{-{\rm i}\delta} \cr} \right ) P^{}_\nu
$ && $\begin{array}{l}
\cos\theta^{}_{12} = |U^{}_{e1}| \\
\tan\theta^{}_{23} = |U^{}_{\tau 1}/U^{}_{\mu 1}| \\
\tan\theta^{\prime}_{23} = |U^{}_{e3}/U^{}_{e 2}| \end{array} $ \\
\hline
{Pattern (3):} ~ $U = R^{}_{23}(\theta^{}_{23}) \otimes
R^{}_{13}(\theta^{}_{13}, \delta) \otimes R^{}_{12}(\theta^{}_{12})
\otimes P^{}_\nu$ && ${\cal J}^{}_l = s^{~}_{12} c^{~}_{12}
s^{}_{23} c^{}_{23} s^{}_{13} c^2_{13} \sin\delta$
\\
$\left ( \matrix{ c^{~}_{12} c^{}_{13}    & s^{~}_{12} c^{}_{13}  &
s^{}_{13} \cr -c^{~}_{12} s^{}_{23} s^{}_{13} - s^{~}_{12} c^{}_{23}
e^{-{\rm i}\delta} & -s^{~}_{12} s^{}_{23} s^{}_{13} + c^{~}_{12}
c^{}_{23} e^{-{\rm i}\delta}      & s^{}_{23} c^{}_{13} \cr
-c^{~}_{12} c^{}_{23} s^{}_{13} + s^{~}_{12} s^{}_{23} e^{-{\rm
i}\delta} & -s^{~}_{12} c^{}_{23} s^{}_{13} - c^{~}_{12} s^{}_{23}
e^{-{\rm i}\delta}      & c^{}_{23} c^{}_{13} \cr} \right ) P^{}_\nu
$ && $\begin{array}{l}
\tan\theta^{}_{12} = |U^{}_{e 2}/U^{}_{e 1}| \\
\tan\theta^{}_{23} = |U^{}_{\mu 3}/U^{}_{\tau 3}| \\
\sin\theta^{}_{13} = |U^{}_{e3}| \end{array} $ \\ \hline
{Pattern (4):} ~ $U = R^{}_{12}(\theta^{}_{12}) \otimes
R^{}_{13}(\theta^{}_{13}, \delta) \otimes
R^{T}_{23}(\theta^{}_{23}) \otimes P^{}_\nu$ && ${\cal J}^{}_l =
s^{~}_{12} c^{~}_{12} s^{}_{23} c^{}_{23} s^{}_{13} c^2_{13}
\sin\delta$
\\
$\left ( \matrix{ c^{~}_{12} c^{}_{13}    & c^{~}_{12} s^{}_{23}
s^{}_{13} + s^{~}_{12} c^{}_{23} e^{-{\rm i}\delta} & c^{~}_{12}
c^{}_{23} s^{}_{13} - s^{~}_{12} s^{}_{23} e^{-{\rm i}\delta} \cr
-s^{~}_{12} c^{}_{13}   & -s^{~}_{12} s^{}_{23} s^{}_{13} +
c^{~}_{12} c^{}_{23} e^{-{\rm i}\delta} & -s^{~}_{12} c^{}_{23}
s^{}_{13} - c^{~}_{12} s^{}_{23} e^{-{\rm i}\delta} \cr -s^{}_{13}
& s^{}_{23} c^{}_{13}   & c^{}_{23} c^{}_{13} \cr} \right ) P^{}_\nu
$ && $\begin{array}{l}
\tan\theta^{}_{12} = |U^{}_{\mu 1}/U^{}_{e1}| \\
\tan\theta^{}_{23} = |U^{}_{\tau 2}/U^{}_{\tau 3}| \\
\sin\theta^{}_{13} = |U^{}_{\tau 1}| \end{array} $ \\ \hline
{Pattern (5):} ~ $U = R^{}_{31}(\theta^{}_{13}) \otimes
R^{}_{12}(\theta^{}_{12}, \delta) \otimes
R^{T}_{13}(\theta^{\prime}_{13}) \otimes P^{}_\nu$ && ${\cal
J}^{}_l = s^2_{12} c^{~}_{12} s^{}_{13} c^{}_{13} s^{\prime}_{13}
c^{\prime}_{13} \sin\delta$
\\
$\left ( \matrix{ c^{~}_{12} c^{}_{13} c^{\prime}_{13} + s^{}_{13}
s^{\prime}_{13} e^{-{\rm i}\delta}  & s^{~}_{12} c^{}_{13} &
-c^{~}_{12} c^{}_{13} s^{\prime}_{13} + s^{}_{13} c^{\prime}_{13}
e^{-{\rm i}\delta} \cr -s^{~}_{12} c^{\prime}_{13}     & c^{~}_{12}
& s^{~}_{12} s^{\prime}_{13} \cr -c^{~}_{12} s^{}_{13}
c^{\prime}_{13} + c^{}_{13} s^{\prime}_{13} e^{-{\rm i}\delta}  &
-s^{~}_{12} s^{}_{13} & c^{~}_{12} s^{}_{13} s^{\prime}_{13} +
c^{}_{13} c^{\prime}_{13} e^{-{\rm i}\delta} \cr} \right ) P^{}_\nu
$ && $\begin{array}{l}
\cos\theta^{}_{12} = |U^{}_{\mu 2}| \\
\tan\theta^{}_{13} = |U^{}_{\tau 2}/U^{}_{e 2}| \\
\tan\theta^{\prime}_{13} = |U^{}_{\mu 3}/U^{}_{\mu 1}| \end{array} $
\\ \hline
{Pattern (6):} ~ $U = R^{}_{12}(\theta^{}_{12}) \otimes
R^{}_{23}(\theta^{}_{23}, \delta) \otimes R^{}_{13}(\theta^{}_{13})
\otimes P^{}_\nu$ && ${\cal J}^{}_l = s^{~}_{12} c^{~}_{12}
s^{}_{23} c^2_{23} s^{}_{13} c^{}_{13} \sin\delta$
\\
$\left ( \matrix{ -s^{~}_{12} s^{}_{23} s^{}_{13} + c^{~}_{12}
c^{}_{13} e^{-{\rm i}\delta} & s^{~}_{12} c^{}_{23}  & s^{~}_{12}
s^{}_{23} c^{}_{13} + c^{~}_{12} s^{}_{13} e^{-{\rm i}\delta} \cr
-c^{~}_{12} s^{}_{23} s^{}_{13} - s^{~}_{12} c^{}_{13} e^{-{\rm
i}\delta} & c^{~}_{12} c^{}_{23}  & c^{~}_{12} s^{}_{23} c^{}_{13} -
s^{~}_{12} s^{}_{13} e^{-{\rm i}\delta} \cr -c^{}_{23} s^{}_{13}
& -s^{}_{23} & c^{}_{23} c^{}_{13} \cr} \right ) P^{}_\nu $ &&
$\begin{array}{l}
\tan\theta^{}_{12} = |U^{}_{e2}/U^{}_{\mu 2}| \\
\sin\theta^{}_{23} = |U^{}_{\tau 2}| \\
\tan\theta^{}_{13} = |U^{}_{\tau 1}/U^{}_{\tau 3}| \end{array} $ \\
\hline
{Pattern (7):} ~ $U = R^{}_{23}(\theta^{}_{23}) \otimes
R^{}_{12}(\theta^{}_{12}, \delta) \otimes
R^{T}_{13}(\theta^{}_{13}) \otimes P^{}_\nu$ && ${\cal J}^{}_l =
s^{~}_{12} c^2_{12} s^{}_{23} c^{}_{23} s^{}_{13} c^{}_{13}
\sin\delta$
\\
$\left ( \matrix{ c^{~}_{12} c^{}_{13}    & s^{~}_{12}    &
-c^{~}_{12} s^{}_{13} \cr -s^{~}_{12} c^{}_{12} c^{}_{13} +
s^{}_{12} s^{}_{13} e^{-{\rm i}\delta} & c^{~}_{12} c^{}_{23} &
s^{~}_{12} c^{}_{23} s^{}_{13} + s^{}_{23} c^{}_{13} e^{-{\rm
i}\delta} \cr s^{~}_{12} s^{}_{23} c^{}_{13} + c^{}_{23} s^{}_{13}
e^{-{\rm i}\delta} & -c^{~}_{12} s^{}_{23} & -s^{~}_{12} s^{}_{23}
s^{}_{13} + c^{}_{23} c^{}_{13} e^{-{\rm i}\delta} \cr} \right )
P^{}_\nu $ && $\begin{array}{l}
\sin\theta^{}_{12} = |U^{}_{e2}| \\
\tan\theta^{}_{23} = |U^{}_{\tau 2}/U^{}_{\mu 2}| \\
\tan\theta^{}_{13} = |U^{}_{e3}/U^{}_{e1}| \end{array} $ \\ \hline
{Pattern (8):} ~ $U = R^{}_{13}(\theta^{}_{13}) \otimes
R^{}_{12}(\theta^{}_{12}, \delta) \otimes R^{}_{23}(\theta^{}_{23})
\otimes P^{}_\nu$ && ${\cal J}^{}_l = s^{~}_{12} c^2_{12} s^{}_{23}
c^{}_{23} s^{}_{13} c^{}_{13} \sin\delta$
\\
$\left ( \matrix{ c^{~}_{12} c^{}_{13}    & s^{~}_{12} c^{}_{23}
c^{}_{13} - s^{}_{23} s^{}_{13} e^{-{\rm i}\delta} & s^{~}_{12}
s^{}_{23} c^{}_{13} + c^{}_{23} s^{}_{13} e^{-{\rm i}\delta} \cr
-s^{~}_{12} & c^{~}_{12} c^{}_{23}  & c^{~}_{12} s^{}_{23} \cr
-c^{~}_{12} s^{}_{13}   & -s^{~}_{12} c^{}_{23} s^{}_{13} -
s^{}_{23} c^{}_{13} e^{-{\rm i}\delta} & -s^{~}_{12} s^{}_{23}
s^{}_{13} + c^{}_{23} c^{}_{13} e^{-{\rm i}\delta} \cr} \right )
P^{}_\nu $ && $\begin{array}{l}
\sin\theta^{}_{12} = |U^{}_{\mu 1}| \\
\tan\theta^{}_{23} = |U^{}_{\mu 3}/U^{}_{\mu 2}| \\
\tan\theta^{}_{13} = |U^{}_{\tau 1}/U^{}_{e1}| \end{array} $ \\
\hline
{Pattern (9):} ~ $U = R^{}_{13}(\theta^{}_{13}) \otimes
R^{}_{23}(\theta^{}_{23}, \delta) \otimes
R^{T}_{12}(\theta^{}_{12}) \otimes P^{}_\nu$ && ${\cal J}^{}_l =
s^{~}_{12} c^{~}_{12} s^{}_{23} c^2_{23} s^{}_{13} c^{}_{13}
\sin\delta$
\\
$\left ( \matrix{ -s^{~}_{12} s^{}_{23} s^{}_{13} + c^{~}_{12}
c^{}_{13} e^{-{\rm i}\delta} & -c^{~}_{12} s^{}_{23} s^{}_{13} -
s^{~}_{12} c^{}_{13} e^{-{\rm i}\delta}      & c^{}_{23} s^{}_{13}
\cr s^{~}_{12} c^{}_{23}    & c^{~}_{12} c^{}_{23}  & s^{}_{23} \cr
-s^{~}_{12} s^{}_{23} c^{}_{13} - c^{~}_{12} s^{}_{13} e^{-{\rm
i}\delta} & -c^{~}_{12} s^{}_{23} c^{}_{13} + s^{~}_{12} s^{}_{13}
e^{-{\rm i}\delta}      & c^{}_{23} c^{}_{13} \cr} \right ) P^{}_\nu
$ && $\begin{array}{l}
\tan\theta^{}_{12} = |U^{}_{\mu 1}/U^{}_{\mu 2}| \\
\sin\theta^{}_{23} = |U^{}_{\mu 3}| \\
\tan\theta^{}_{13} = |U^{}_{e3}/U^{}_{\tau 3}| \end{array} $
\end{tabular}
\end{center}
\end{table}

\end{document}